\documentclass[11pt,a4paper]{article}

\title{Co-impact: Crowding Effects in Institutional Trading Activity}
\author{Fr\'ed\'eric Bucci£$^1$, Iacopo Mastromatteo$^2$, Zolt\'an Eisler$^2$, \\
Fabrizio Lillo$^3$, Jean-Philippe Bouchaud$^{2,4}$ and Charles-Albert Lehalle$^{2,4}$\\}
\date{%
$^1$ \textit{Scuola Normale Superiore, Piazza dei Cavalieri 7, 56126 Pisa, Italy} \\%
$^2$ \textit{Capital Fund Management, 23-25, Rue de l'Universit\'e 75007 Paris, France}  \\%
$^3$ \textit{Department of Mathematics, University of Bologna,  Piazza di Porta San Donato 5, 40126 Bologna, Italy}\\[2ex]%
$^4$ \textit{CFM-Imperial Institute of Quantitative Finance, Department of Mathematics, Imperial College, 180 Queen's Gate, London SW7 2RH} \\[2ex]%
\today}

\usepackage{amssymb}
\usepackage{graphicx}
\usepackage{amsmath}
\usepackage{amsfonts}
\usepackage{geometry}
\usepackage{bm}
\usepackage{breqn}
\usepackage{textcomp}
\usepackage{xcolor}
\usepackage{subfig}
\usepackage{amsmath}
\usepackage{amssymb}
\usepackage{geometry}
\usepackage{mathtools}
\usepackage{amsthm}
\usepackage{enumerate}
\theoremstyle{definition}
\newtheorem{definition}{Remark}
\newtheorem{def1}{Step}

\newcommand{\dd}{\mathrm{d}}
\newcommand{\avg}{\mathbb{E}}
\newcommand{\var}{\mathbb{V}}
\newcommand{\corr}{\mathbb{C}}
\newcommand{\imp}{\mathcal{I}}
\newcommand{\sign}{{\mathrm{sign}}}

\newcommand{\et}{\tilde{\epsilon}}

\newcommand{\bphi}{{\bm \varphi}}

\newcommand{\volfl}{\Sigma}

\newcommand{\orange}[1]{{\color{orange} FABRIZIO: #1}}

\begin{document}

\maketitle

\abstract{This paper is devoted to the important  yet unexplored subject of \emph{crowding} effects on market impact, that we call ``co-impact''. Our
analysis is based on a large database of metaorders by institutional investors in the U.S. equity market. We find that the market chiefly reacts to
the net order flow of ongoing metaorders, without individually distinguishing them. The joint co-impact of multiple contemporaneous metaorders depends
on the total number of metaorders and their mutual sign correlation. Using a simple heuristic model calibrated on data, we reproduce very well the
different regimes of the empirical market impact curves as a function of volume fraction $\phi$: square-root for large $\phi$, linear for intermediate $\phi$, and a finite
intercept $I_0$ when $\phi \to 0$. The value of $I_0$ grows with the sign correlation coefficient. Our study sheds light on an apparent paradox: How
can a non-linear impact law survive in the presence of a large number of simultaneously executed metaorders? }

\tableofcontents

\section{Introduction}

The market impact of trades, i.e. the change in price conditioned on signed trade size, is a key quantity characterizing market liquidity and price
dynamics \cite{bouchaud2009markets,book2}. 
Besides being of paramount interest for any economic theory of price formation,
impact is a major source of transaction costs, which often makes the difference between a trading
strategy that is profitable, and one that is not. Hence the interest in this topic is not purely academic in nature.

One of the most surprising empirical finding in the last 25 years is the fact that the impact of a so-called ``metaorder'' of total size $Q$, executed incrementally over time, increases
approximately as the {\it square root} of $Q$, and not linearly in $Q$, as one may have naively expected and as indeed predicted by the now-classic
Kyle model \cite{kyle1985continuous}. Since impact is non-additive, a natural question concerns the {\it interaction} of different metaorders executed
simultaneously -- possibly with different signs and sizes. In particular, one may wonder whether the simultaneous impact of different metaorders could
substantially alter the square root law; or conversely whether the square root law might itself result from the interaction of different metaorders. 

Metaorder information is, however, not publicly available, and earlier analyses were mostly based on (often proprietary) data from single financial 
institutions. These
studies give little insight about effects due to the simultaneous execution of metaorders from different investors, which we will call {\it co-impact} hereafter. 
Indeed, even if investors individually decide about their metaorders, they might do so based on the same trading signal. Prices can thus be affected by emergent effects such
as crowding. What is the right way to model the total market impact of simultaneous metaorders on the same asset on the same day? In order to answer
this question, we will use a rich dataset concerning the execution of metaorders issued by a heterogeneous set of investors.

The paper is organized as follows. In Sec.~\ref{sec:Ancerno} we introduce the Ancerno dataset we used empirically.
In Sec.~\ref{sec:sqrt} we discuss the limits of the validity of the square root law on the daily level. In Sec.~\ref{sec:coimpact} we find that the market impact of
simultaneous daily metaorders is proportional to the square root of their net order flow. This means that the market does not 
distinguish the different individual metaorders. We then construct a theoretical framework to understand the impact of correlated metaorders 
in Sec.~\ref{sec:stats}. This allows us to understand when a single asset manager will observe a square-root impact law, and when {\it crowding
effects} will 
lead to deviations from such a behaviour. We also compare in Sec.~\ref{sec:stats} the results of our simple mathematical model with empirical data, with very satisfactory results. Sec.~\ref{sec:conclusions} concludes.

\section{The Ancerno Database}
\label{sec:Ancerno}
Our analysis relies on a database made available by {\it Ancerno}, a leading
transaction-cost analysis provider\footnote{Ancerno Ltd. (formerly the Abel Noser Corporation) is a widely recognised consulting
firm that works with institutional investors to monitor their equity trading
costs. Its clients include many pension funds and asset managers. Previous academic studies that use Ancerno data include
\cite{zarinelli2015beyond,puckett2008short,goldstein2009brokerage,chemmanur2009role,jame2010organizational,goldstein2011purchasing,puckett2011interim,
busse2012buy}. See {\tt www.ancerno.com} for details.}. The unique advantage of working with such
institutional data is that one can simultaneously analyze the trading of many investors. The main caveat though is that one has
little knowledge about the motives and style behind the observed portfolio transitions. For example a given metaorder can be part of a longer execution over multiple
days. Another possibility is that the final investor may decide to stop a metaorder execution midway if prices move unfavourably. Such effects can potentially bias our
results, but we believe that they do not change the qualitative conclusions below.

In the following we will define as a metaorder a series of jointly reported executions performed by a single investor, through a single broker within
a single day, on a given stock and in a given direction (buy/sell). However, contrarily to the version of the database used in
Ref.~\cite{zarinelli2015beyond},
available labels do not allow us to relate different metaorders executed on behalf of 
the same final investor by the same or different brokers during the same day. These should ideally be counted as a single
metaorder. We will comment later on the biases induced by such a lack of information. Thus each metaorder is characterized by a
broker label, the stock symbol, the total volume of the metaorder $|Q|$ and its sign $\epsilon=\pm 1$, and the start time $t_s$ and the end time $t_e$
of the execution.

\begin{figure}[th]
\begin{center}
\includegraphics[width=1.0\textwidth]{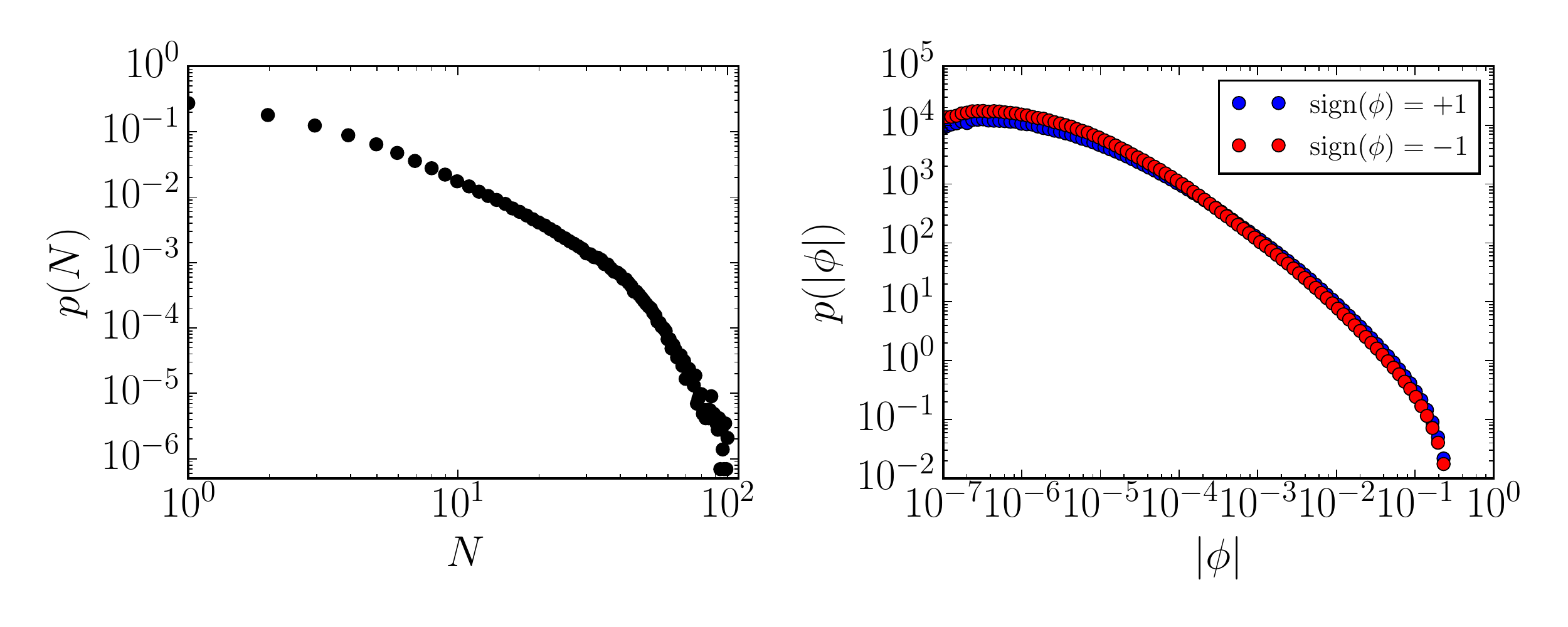}
\caption{(Left panel) Empirical probability distribution of the number $N$ of daily metaorders per asset.
(Right panel) Empirical probability distribution of the absolute value of the volume fraction $\phi$ per metaorder, separately for signs $\pm 1$,
i.e. buy/sell.}
\label{figure7}
\end{center}
\end{figure}

Our dataset includes the period January  2007~--~June 2010 for a total of 880 trading days. Following the  procedure introduced in
Ref.~\cite{zarinelli2015beyond} we use the following
filters to remove possibly erroneous data:
\begin{itemize}
\item \textbf{Filter 1}: We select the stocks which belong to the Russell 3000 index discarding metaorders executed on highly illiquid stocks.
\item \textbf{Filter 2}: We select metaorders ending before 4:01 p.m.
\item \textbf{Filter 3}: We select metaorders whose duration $D= t_e - t_s$ is longer than 2 mins.
\item \textbf{Filter 4}: We select metaorders whose participation rate (the
ratio between their quantity and the volume traded by the market between $t_s$ and $t_e$) is smaller than 30\%.
\end{itemize}
Finally we retain around 7.7 million metaorders distributed quite uniformly in time and across market capitalizations. These filtered metaorders represent around the
5\% of the total reported market volume independently of the year and of the stock capitalization.\footnote{Without the above filters, this number
would rise to about $10 \%$.}
The statistical properties of the metaorders, in terms of volume fraction, duration, etc., are broadly in line with Ref.~\cite{zarinelli2015beyond} 
even though their data was aggregated at the level of brokers -- for more details see Appendix~\ref{app:metastat}. A particularly important statistic for the following analyses 
is the number $N$ of
simultaneous metaorders in the database, executed on the same stock during the same day. The probability distribution $p(N)$ is shown in the left panel Fig.~\ref{figure7}, indicating that $N$ is broadly distributed with an average close to $5$.

The right panel of  Fig.~\ref{figure7} shows the probability distribution
of the absolute value of the volume fraction $\phi$ of the metaorders. This variable plays a key role in the following and is defined as 
$$\phi:=Q/V$$
where $V$ is the total volume traded during that day. The figure shows that the volume fraction distribution is independent of the metaorder side
($\textrm{sign}(\phi)=\pm 1$, i.e.\ buy or sell) and is also very broad.

\section{The Square Root Law and its Domain of Validity}
\label{sec:sqrt}

We will quantify market impact in terms of the rescaled log-price $s=\log (S)/\sigma$, where $S$ is the market mid-price, which we normalize by the
daily
volatility of the asset defined as $\sigma=(S_{\mathrm{high}}-S_{\mathrm{low}}) / S_{\mathrm{open}}$ based on
the daily high, low and open prices. In this paper we will define impact as the expected change of $s$ between the open and the close of the day. This
choice will avoid an elaborate analysis of when precisely each metaorder starts and ends, how they overlap and which reference prices to take in each
case. When a metaorder of total volume $Q$ is executed, its impact will be defined as
\begin{equation}
I(\phi) := \avg[s_{\mathrm{close}}-s_{\mathrm{open}}|\phi],
\label{IQ1}
\end{equation}
for a given metaorder signed volume fraction $\phi$.

Empirically, impact is found to be an odd function of $\phi$, displaying a concave behavior in $|\phi|$. It is well described by the square root
law~\cite{toth2011anomalous,mastromatteo2014agent,torre1998market,almgren2005direct,engle2006measuring,brokmann2015slow,bacry2015market,
moro2009market}
\begin{equation}
I(\phi)=Y \times \phi^{\bullet\delta},
\label{IQ}
\end{equation}
where here and throughout the paper we will denote the sign-power operation by $x^{\bullet \delta}:=\sign(x) \times |x|^\delta$.
The dimensionless coefficient $Y$ (called the \emph{Y-ratio}) is of order unity and the exponent
$\delta$ is in the range 0.4--0.7. It is interesting to note that in Eq.~\eqref{IQ} only the  volume fraction $\phi$ matters, the
time taken to complete execution or the presence of other active metaorders is not directly relevant (remember that the volatility 
of the instrument has been subsumed in the definition of the rescaled price $s$). This formula is surprisingly universal across financial products, market venues, 
time periods and the strategies used for execution.

\begin{figure}[th]
\begin{center}
\includegraphics[width=0.5\textwidth]{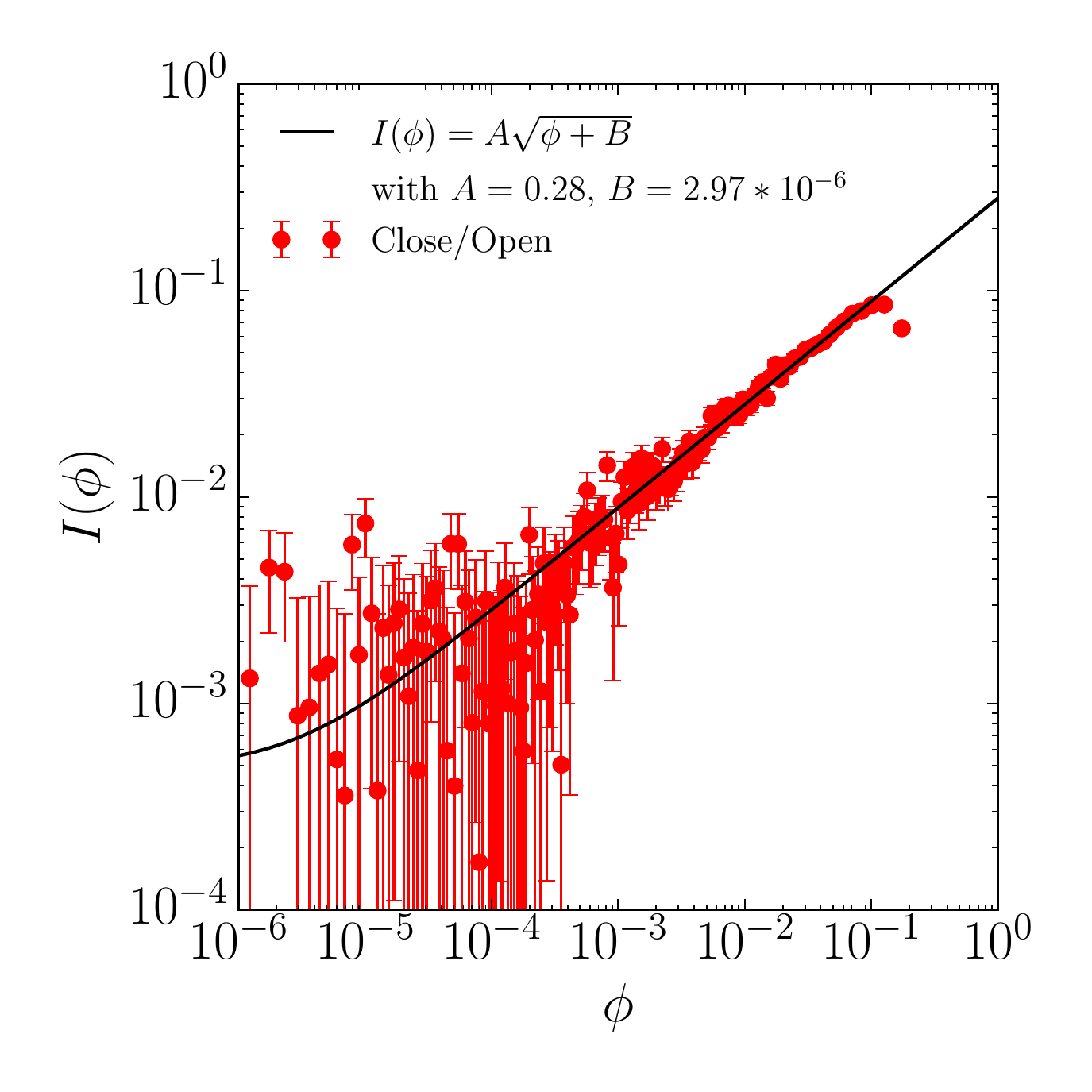}
\caption{
Market impact curve $I(\phi) = \avg[s_{\mathrm{close}}-s_{\mathrm{open}}|\phi]$ as a function of the
metaorder size ratio $\phi=Q/V$ computed using the filtered metaorders from the Ancerno dataset in the period from January 2007~--~June
2010. We also show the simple fit $I(\phi)=A \sqrt{\phi+B}$, which captures some -- but not all -- of the discrepancy with the square root law at
small $\phi$.}
\label{fig:IxpPXCO}
\end{center}
\end{figure}

We first check this empirical result on our dataset. 
In Fig.~\ref{fig:IxpPXCO} we show the market impact curve obtained by dividing the data into evenly populated bins according to the  volume fraction $\phi$ and computing the conditional expectation of impact for each bin. Here and in the following, error bars are determined as standard errors.
Note that in all the following empirical plots  the price impact curves are normalized by their \emph{Y-ratio} and we will abuse the notation $I(\phi)$ in order to denote the symmetrized measure $I(\phi) = \avg[\epsilon(s_{\mathrm{close}}-s_{\mathrm{open}})| |\phi|]$, with $\epsilon = \sign(\phi)$, due to the antisymmetric nature of $I(\phi)$.

While the square root law holds relatively well when $10^{-3} \lesssim \phi \lesssim 10^{-1}$, three other regimes seem to be present:
\begin{enumerate}
\item For very small volume fractions up to $\phi \lesssim 10^{-4}$, impact appears to saturate to a finite, positive value.
\item In the intermediate regime $10^{-4} \lesssim \phi \lesssim 10^{-3}$, impact is closer to a linear function, although the data is very noisy. 
\item In the large $\phi$ regime $\phi \gtrsim 10^{-1}$, impact seems to saturate, or even to decrease with increasing $\phi$. 
\end{enumerate}

These results are robust across time periods and market capitalizations, and consistent with Ref.~\cite{zarinelli2015beyond}, where regimes 2. and 3. were also clearly observed. 
In the following, we will
discard altogether the last, large $\phi$ regime, which is most likely affected by conditioning effects 
(for example buying more when the
price moves down and less when the price moves up). We will on the other hand seek to understand the other three regimes within a consistent mathematical 
framework.

Intuitively, the breakdown of the square root law for small $\phi$ comes from the fact that the signs of the metaorders in our dataset are
correlated -- particularly so 
because some metaorders are originating from the same final investor.  
Let us illustrate the effect of correlations on a simplistic example: Imagine that simultaneously to the considered buy metaorder (with volume
fraction $\phi > 0$), another metaorder with the 
same sign and volume fraction $\phi_m > 0$ is also traded. Assuming that the square root law applies for the combined metaorder (a hypothesis that we
will confirm on data), the
observed impact should read
\begin{equation}
I(\phi+\phi_m)=Y \times \sqrt{\phi+\phi_m}.
\label{IQ2}
\end{equation}
This tends to the value $Y \sqrt{\phi_m}$ when $\phi \to 0$, behaves linearly when $\phi \ll \phi_m$ and as a square root when $\phi \gg \phi_m$. We show in
Fig.~\ref{fig:IxpPXCO} that this simple fit captures some, but not all, of the discrepancy with the square root law at small $\phi$. In
particular the intermediate linear region is not well accounted for. We will develop in the following a mathematical model that reproduces all
these effects. 

\begin{figure}
\begin{center}
\includegraphics[width=0.5\linewidth]{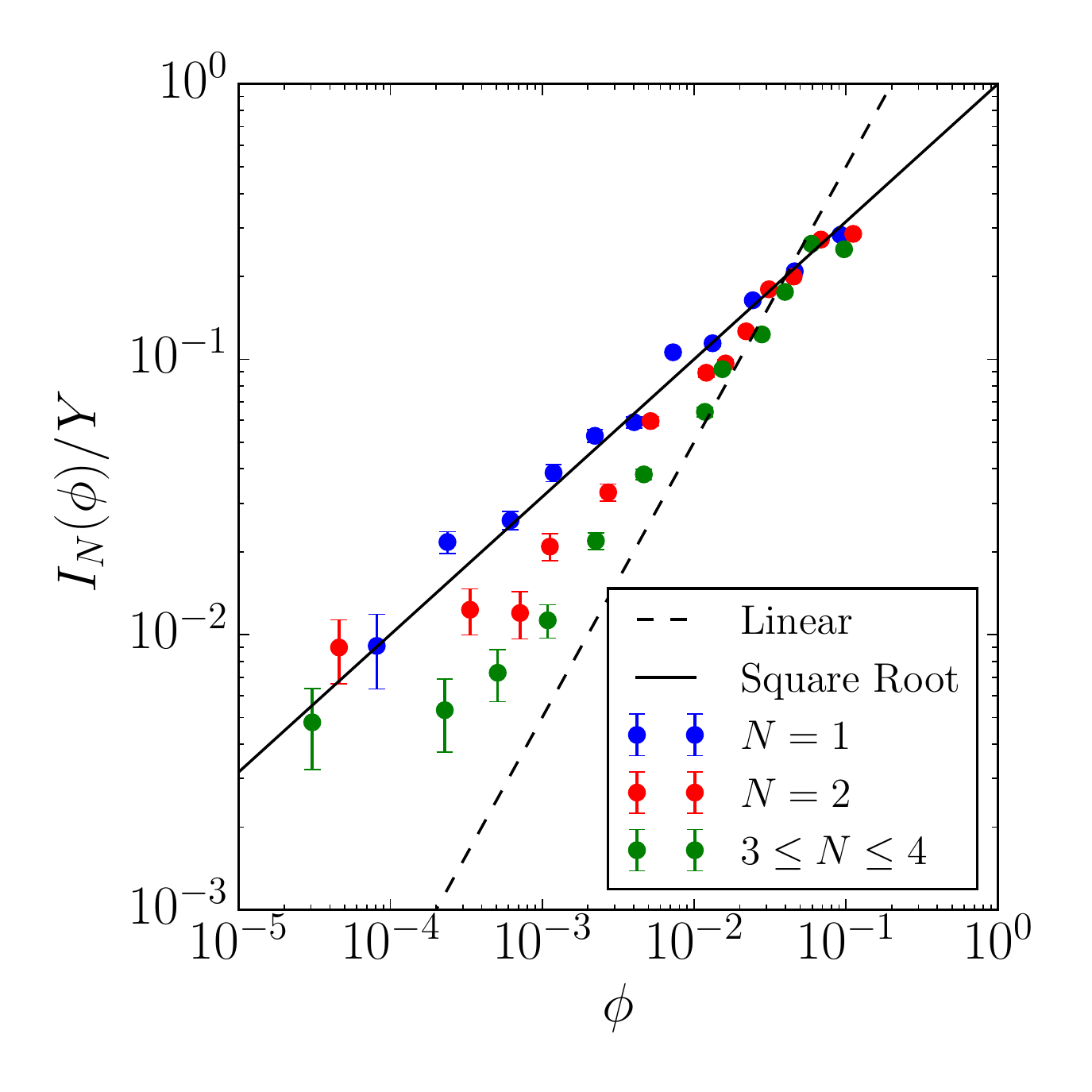}
\caption{Empirical evidence on the effect of the number of metaorders $N$ on the daily price impact curves $I_N(\phi)$ normalized by the prefactor
\emph{Y-ratio}: significant departures from the square root law can be observed for small $\phi$  increasing the number $N$ of daily metaorders per asset. }
\label{fig:empiricalNrho}
\end{center}
\end{figure}

A way to minimize the effect of correlations is to restrict to days/assets where there is a unique metaorder in the dataset ($N=1$). As shown in
Fig.~\ref{fig:empiricalNrho}, impact in this case is almost perfectly fitted by a square root law. Fig.~\ref{fig:empiricalNrho}, also shows that as
$N$ increases, significant departures from the square root law can be observed for small $\phi$, as suggested by our simple model Eq. \eqref{IQ2}. An
alert reader may however object that the Ancerno database represents a small fraction ($\sim 5 \%$) of the total volume. Even when 
a single metaorder is reported, many other metaorders are likely to be simultaneously present in the market. So why does one observe a square root law
at all, even for single metaorders? The solution to precisely this paradox is one of the main messages of our paper.   

\section{How Do Impacts Add Up?}
\label{sec:coimpact}

In the previous section we showed that the number of metaorders in the market strongly influences how price impact behaves, but we have yet to provide insight into why
this is the case. As a first step, we want to determine an explicit functional form of the aggregated market impact of $N$ simultaneous metaorders. As we have emphasized, impact
is non-linear, so aggregation is a priori non-trivial. Should one add the square root impact of each metaorder, or should one first add the signed
volume fractions before taking the
square root? Since orders are anonymous and indistinguishable, the second procedure looks more plausible. This is what we test now. Consider the
average aggregate impact conditioned to the co-execution of $N$ metaorders:
\begin{equation}
  \imp(\bphi_N) = \avg[s_{\mathrm{close}} - s_{\mathrm{open}} | \bphi_N],
\end{equation}
where $\bphi_N:=(\phi_1,\cdots,\phi_N)$. We make the following parametric ansatz for this quantity: 
\begin{equation}
\imp(\bphi_N)=Y\times \left( \sum_{i=1}^N \phi_i^{\bullet \alpha} \right)^{\bullet \delta/\alpha},
\label{eq:formula}
\end{equation}
where, again, $x^{\bullet \alpha}$ is the signed power of $x$. By construction this formula is invariant under the permutation of metaorders, as
it should be since they are indistiguishable.
$Y$ and $\delta$ set, respectively, the scale and the exponent of the impact function. The free parameter
$\alpha$ interpolates between the case when impacts add up ($\alpha=\delta$) and when only the net traded volume is relevant ($\alpha = 1$).

\begin{figure}[th!]
\begin{center}
\includegraphics[width=0.52\textwidth]{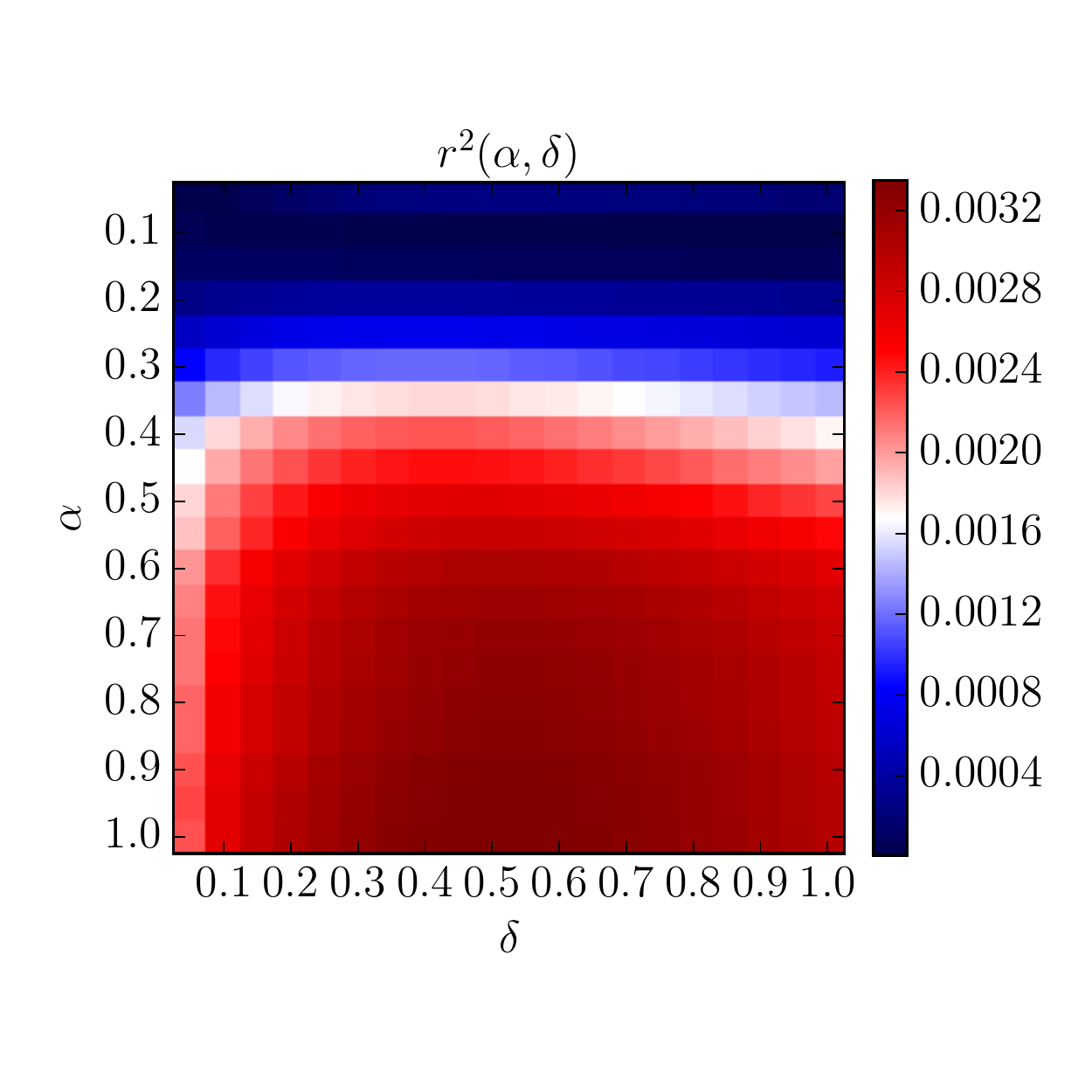}
\caption{The computed coefficient of determination
$r^2(\alpha,\delta)$ of the least squares regressions of Eq.~\eqref{eq:formula} for a grid of $(\alpha,\delta)$ pairs. The coefficient of
determination is maximized ($r^2=0.0035$) in the vicinity of the point
$\alpha=1.0$ and $\delta=0.5$.}
\label{fig:heatmap}
\end{center}
\end{figure}

Fig.~\ref{fig:heatmap} shows the quality of the fit obtained by least squares regressions of Eq.~\eqref{eq:formula} for a grid
of $(\alpha,\delta)$ pairs. We find that the coefficient of determination $r^2(\alpha, \delta)$ of the fit is maximized close to the point $\alpha=1.0$ and $\delta=0.5$, 
which suggests that the aggregated price impact $\imp(\bphi_N)$ of $N$ metaorders at the daily scale only depends on the total net order flow, i.e.\footnote{We have in fact tested that the assumption $\alpha=1$ is also favoured for a general, non parametric shape for the impact function $\mathcal{I}(\Phi)$.} 
\begin{equation}
  \imp(\bphi_N) \approx Y \times \Phi^{\bullet 1/2},
  \label{eq:formula2}
\end{equation}
where $\Phi = \sum_{i=1}^N \phi_i$. In other words, the market only reacts to the net order flow, not to the way in which this order flow is
distributed across investors. 

\begin{figure}[th!]
\begin{center}
\includegraphics[width=0.5\textwidth]{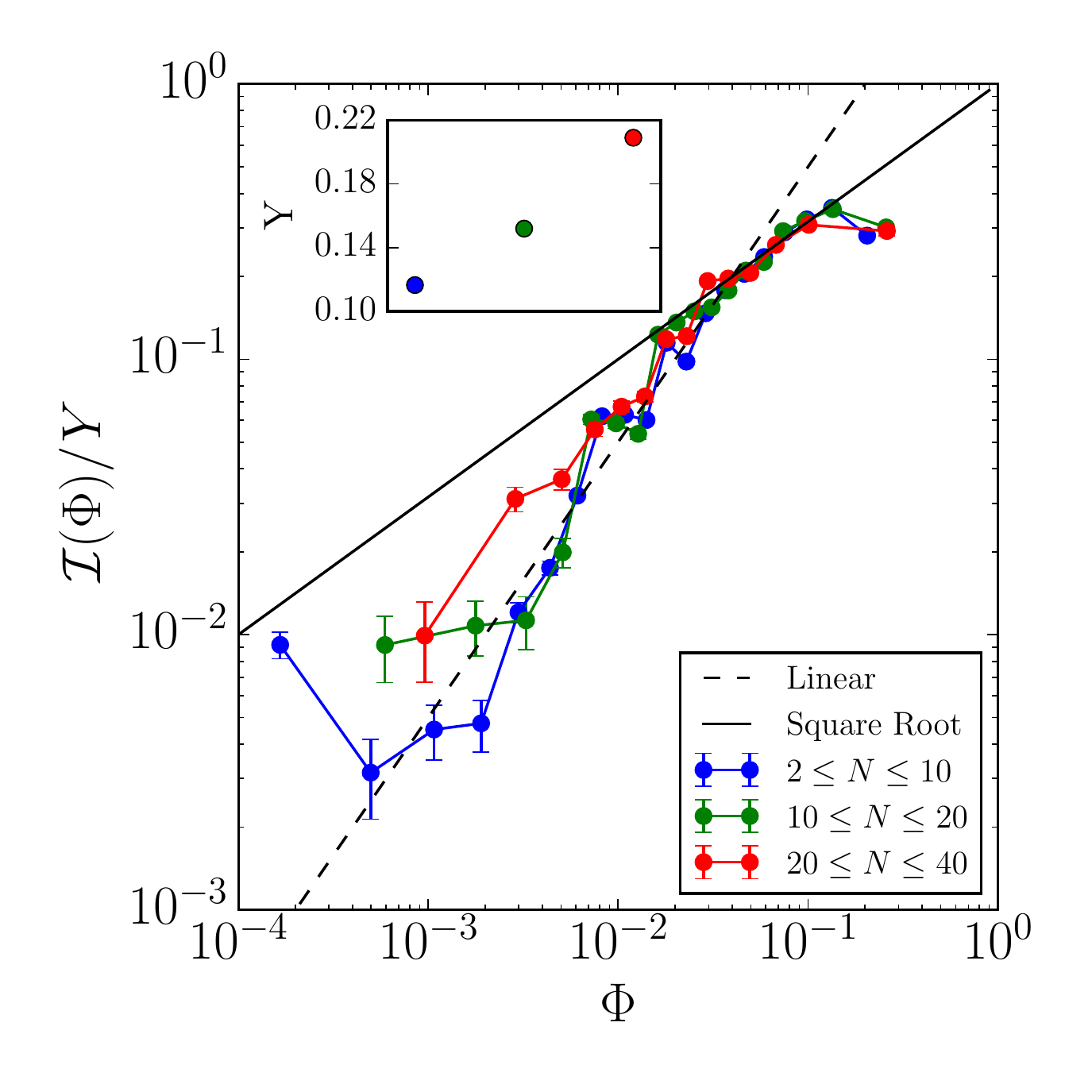}
\caption{Global market impact $\mathcal{I}(\Phi)=\avg[\mathcal{I}(\bphi_N)|\sum_{i=1}^N\phi_i=\Phi]$ normalized by the prefactor $Y$ as a function of
the net order flow $\Phi$ for various buckets in  $N$ and symmetrized as explained in Sec.~\ref{sec:sqrt}. The inset shows the normalization $Y$ for
the different curves.}
\label{fig:Itilde}
\end{center}
\end{figure}

One can now plot $\imp(\Phi)$ as a function of $\Phi$ for various $N$, see Fig.~\ref{fig:Itilde}. One clearly sees that provided $\Phi$ is not too
small, impact is independent of $N$ 
and crosses over from linear to square root behavior as $\Phi$ increases. The linear behavior is in fact more pronounced at smaller
values of $N$.  We now turn to a theoretical analysis that will allow us to quantify more precisely the co-impact problem, and how the square root law
can survive at large $N$.

\section{Correlated Metaorders and Co-Impact}
\label{sec:stats}

\subsection{The Mathematical Problem}
Even if an asset manager knows the average impact formula Eq. \eqref{eq:formula2}, this may not be sufficient to estimate  his actual impact which
depends strongly on the presence of other contemporaneous metaorders.

Suppose the manager $k$ wants to execute a volume fraction $\phi_k=\phi$.
If all the other $N-1$ metaorders were known, the daily price impact would be given by the global impact function
$\imp(\Phi)$ determined in the previous section, with $\Phi=\phi_k+\sum_{i\neq k}^N\phi_i$.

However, this information is obviously not available to
the manager $k$. His best estimate of the average impact given $N$ is the conditional expectation
\begin{equation}
I_N(\phi) = \avg[\imp(\Phi)|\phi_k=\phi] =
\mathbb{E}\Big[\imp\Big(\phi_k+\sum_{i\neq k}^N \phi_i\Big)\Big|\phi_k=\phi\Big]
\label{eq:Iphi}
\end{equation}
over the conditional distribution $P(\bphi_N|\phi_k=\phi)$ of the metaorders. Since the number of metaorders is in general not known either, the
expected individual market impact is given by 
\begin{eqnarray}
\label{eq:Iphitot}
I(\phi) = Y \times \sum_N p(N) \int \dd \phi_1 \dots \dd \phi_N  P(\bphi_N| \phi_k = \phi)  \Big(\phi_k+\sum_{i\neq k}^N\phi_i \Big)^{\bullet 1/2},
\end{eqnarray}
where we have used Eq.~\eqref{eq:formula2}.
In such a way to compute $I_N(\phi)$ and $I(\phi)$ we need to know the joint probability density function $P(\bphi_N):= P(\phi_1,\dots, \phi_N)$, which is
in general a complicated and high-dimensional object. Then to create a tractable model that can be calibrated on data, we must make
some reasonable assumption on the dependence structure of the $\phi_i$. In the next subsection we investigate the simple case where the $\phi_i$ are all
independent, and then turn to an empirical characterization of the correlations between metaorders. We finally provide the results of our empirically
inspired model and compare them with the empirical market impact curves.

\subsection{Independent Metaorders}
\label{gauss}

\begin{figure}
\begin{center}
\centering
\includegraphics[width=1.0\textwidth]{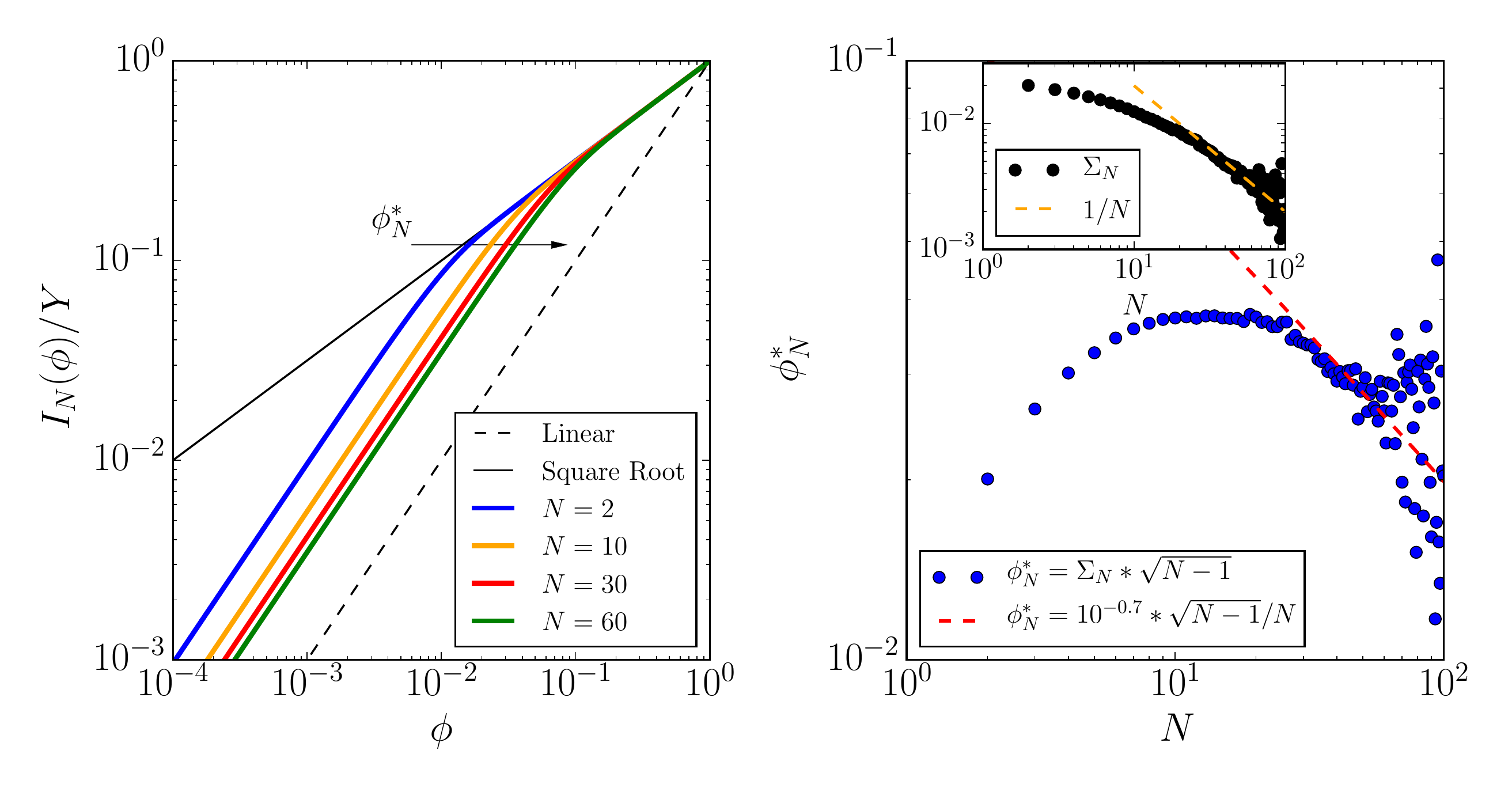}
\caption{(Left panel) Market impact curves $I_N(\phi)/Y$ for i.i.d.\ Gaussian metaorders for different $N \in \{2, 10, 30, 60\}$ and {\it fixed} $\Sigma_N = \Sigma = 0.8\%$ computed from empirical data. 
The transition from the square root to the linear regime takes place for
$\phi^*_N \simeq \Sigma \sqrt{N-1}$. (Right panel) Empirically estimated $\phi_N^*$ as a function of the number $N$ of daily metaorders per asset. This is obtained by computing the empirical order size's standard deviation  $\Sigma_N=\sqrt{\mathbb{V}[\phi|N]}$ conditioned to $N$.
The dashed red line shows the case in which $\Sigma_N\sim 1/N$ for each $N$: in the inset we report the standard deviation $\Sigma_N$ of the order size $\phi$ as a function of the number $N$ of daily metaorders per asset.}
\label{fig:nonintera}
\end{center}
\end{figure}

The simplest assumption about the form of $P(\bphi_N)$ is that metaorder volumes are i.i.d., meaning
\begin{equation}
P(\bphi_N)=\prod_{i=1}^N p(\phi_i).
\end{equation}
Assuming for simplicity that each $\phi_i$ is a Gaussian random variable with zero mean and variance $\volfl_N^2$, where the lower index indicates
an explicit dependence on $N$. Thus $N-1$ simultaneous metaorders generate a Gaussian noise contribution of amplitude 
$\volfl_N\sqrt{N-1}$ on top of $\phi_k=\phi$. In Appendix~\ref{app:calciidgauss} we show analytically that: 
\begin{itemize}
\item For small metaorders the noise term dominates, leading to
$$I_N(\phi) \propto \phi \quad \mathrm{when} \quad \phi \ll \phi^*_N:=\volfl_N\sqrt{N-1}.$$
\item For large metaorders the $N-1$ other simultaneous metaorders can be neglected and thus
$$I_N(\phi) \propto \sqrt{\phi} \quad \mathrm{when} \quad \phi \gg \phi^*_N.$$
\end{itemize}
In Appendix~\ref{app:calciidnongauss} we show that the above results remain valid in the limit of large $N$ independently of the shape of the volume distribution provided its variance is finite.

The full analytical solutions for different $N$ values, but {\it fixed} $\volfl_N = \Sigma$, are shown in the left panel of Fig.~\ref{fig:nonintera}. One
clearly sees the cross-over from a linear behavior at small $\phi$ 
to a square root at larger $\phi$. However, interestingly, one expects $\volfl_N$ to {\it decrease} with $N$, simply because as the number of
metaorders increases, the volume fraction represented by each of 
them must decrease\footnote{For example, the variance of a flat Dirichlet random variable $(X_1,...,X_N)\sim \text{Dir}(N)$ describing fractions is $\mathbb{V}[X_i|N]=(N-1)/(N^2(N+1))\sim N^{-2}$.}. As shown in the right panel of Fig.~\ref{fig:nonintera}, this is the case empirically since for $N \gtrsim 10$, $\volfl_N$ indeed decays as $N^{-1}$ (as also suggested by Fig.~\ref{fig:Zeta} below). Hence, for large $N$, the crossover value
$\phi^*_N$ {\it decreases} with $N$ as $N^{-1/2}$. This explains why the square root law can at all be observed when a large number of metaorders are
present. If these metaorders are independent, their net impact on the
price averages out, leaving the considered metaorder as if it was alone in a random flow, as assumed in theoretical models
\cite{toth2011anomalous,mastromatteo2014agent}. We now turn to the effect of correlations between metaorders.

\subsection{Metaorder Correlations}

\begin{figure}
\begin{center}
\includegraphics[width=0.5\textwidth]{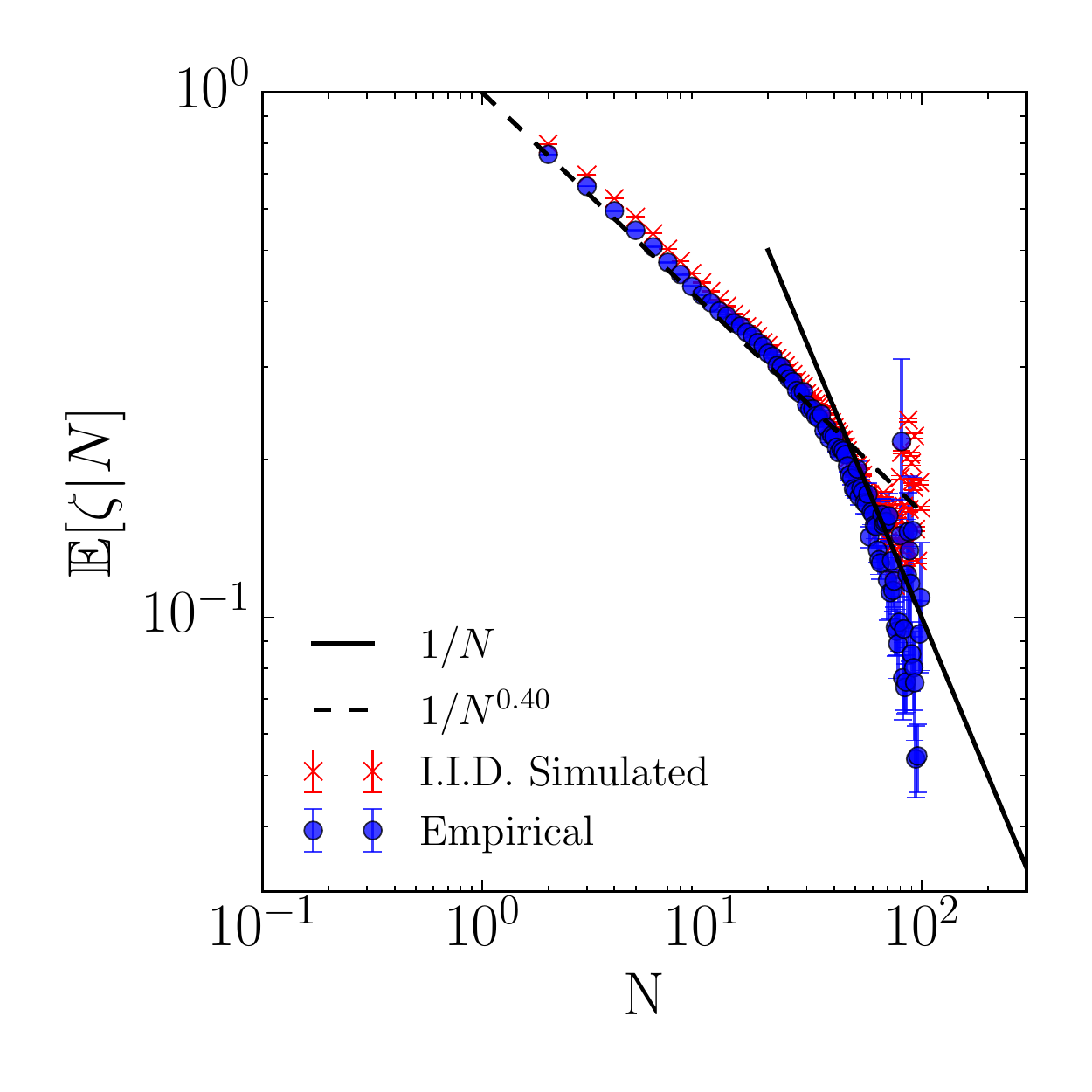}
\caption{Average value of the Herfindahl index (or ``inverse participation ratio'') $\avg[\zeta|N]$ as a function of the number $N$ of daily metaorders per asset computed from  the empirical data  (blue dot symbols) and simulating i.i.d.\ $|\phi_i|$  (red star symbols) extracted from the empirical distribution illustrated in the right panel of Fig.~\ref{figure7}.}
\label{fig:Zeta}
\end{center}
\end{figure}

In order to build a sensible model of $P(\bphi_N)$ we consider separately the size distribution and the size cross-correlations. From the
right panel of Fig.~\ref{figure7} we observe that the marginals $p(\phi_i)$ are to a good approximation independent of the direction,
buy or sell, and moderately fat tailed. The latter observation suggests that the total net order flow $\Phi = \sum_{i=1}^N \phi_i$ is {\it not} dominated by a single
metaorder. A way to
quantify this is through the Herfindahl index (or ``inverse participation ratio'') $\zeta$, defined as:
\begin{equation}
\zeta:= \frac{\sum_{i=1}^N \phi_i^2}{\left(\sum_{i=1}^N |\phi_i|\right)^2}.
\end{equation}
This quantity is of order $1/N$ if all metaorders are of comparable size, and of order $1$ if one metaorder dominates. In Fig.~\ref{fig:Zeta}  we show
the dependence of $\mathbb{E}[\zeta|N]$ as a function of $N$, which clearly decays with $N$. It also compares very well to the result obtained
assuming the absolute volume fractions $|\phi_i|$ to be independent, identically distributed variables, drawn according to the empirical distribution
shown in Fig.~\ref{figure7}.
We therefore conclude that (a) metaorders in the Ancerno database are typically of comparable relative sizes $\phi$ and (b) absolute volume correlations do not play a major role, and we will neglect them henceforth.

\begin{figure}[t!]
\begin{center}
\includegraphics[width=0.5\textwidth]{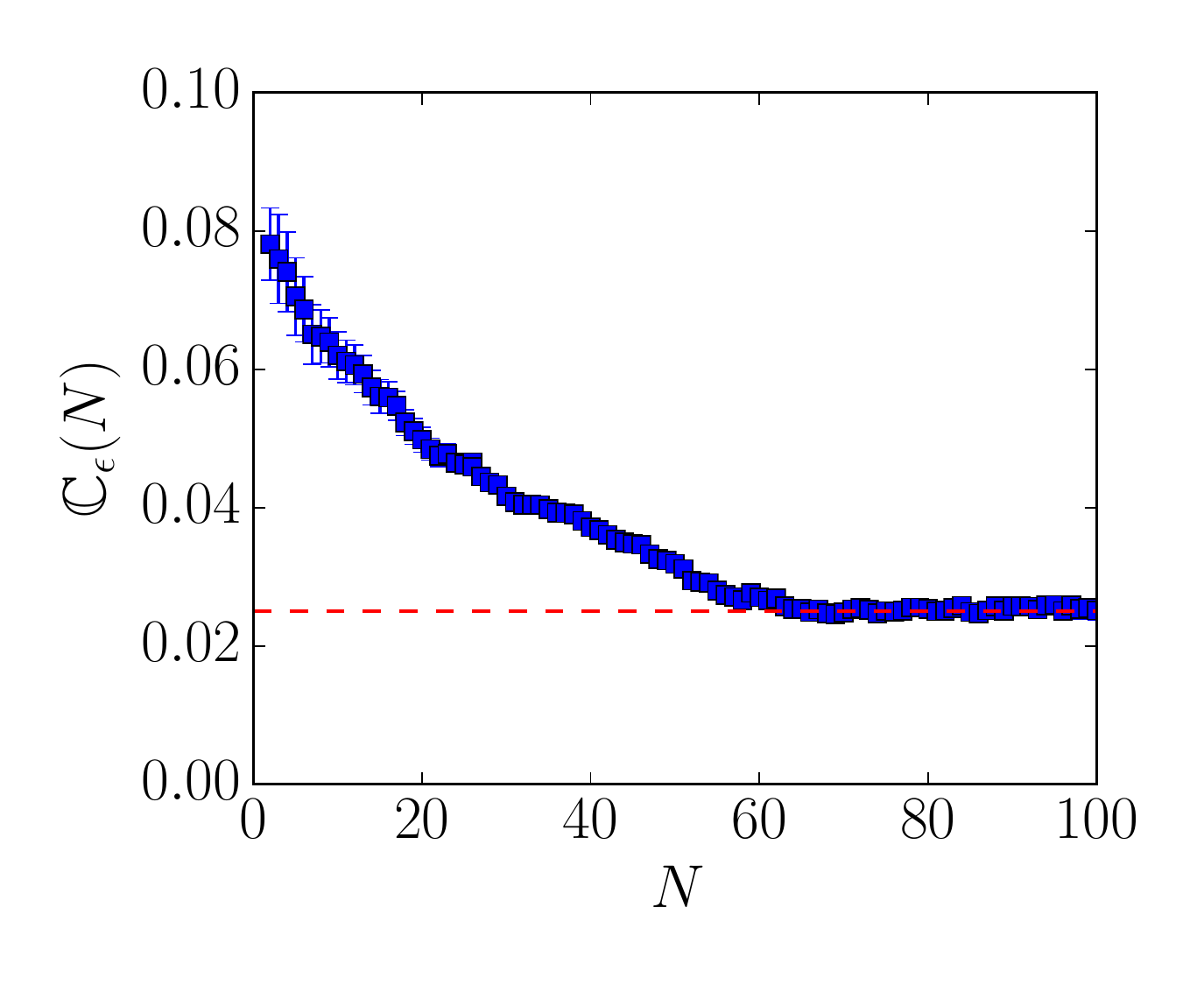}
\caption{Empirical average sign correlation $\corr_{\epsilon}(N)$ as a function of the number  $N$ of daily metaorders: the dashed red line
represents the plateau value $\corr_{\epsilon} \approx 0.025$ at large $N$, which, we believe, is a reasonable proxy for the correlation of orders
submitted by different asset managers.}
\label{figure9}
\end{center}
\end{figure}

Sign correlations, on the other hand, \emph{do} play an important role in determining the impact of simultaneous metaorders. The empirical average
sign correlation of metaorders simultaneously executed on the same asset 
is defined as
\begin{equation}
  \label{corr1}
  \corr_{\epsilon}(N) :=  \frac{\mathbb{E}[\epsilon_i\epsilon_j|N]- \mathbb{E}[\epsilon_i|N]^2}{\mathbb{E}[\epsilon_i^2|N]-\mathbb{E}[\epsilon_i|N]^2} \, ,
\end{equation}
where $\avg[\cdots|N]$ is the average over all days and assets such that exactly $N$ metaorders were executed. Fig.~\ref{figure9} shows the dependence
of $\corr_{\epsilon}$ on $N$. We clearly see that on average the daily metaorders executed on the same asset are positively correlated. Furthermore,
$\corr_{\epsilon}(N)$ is seen to decrease as $N$ increases. This is likely due to the fact that there are multiple concurrent metaorders submitted by
the same manager, an effect that becomes less prominent as $N$ increases. The plateau value $\corr_{\epsilon} \approx 0.025$ at large $N$ is, we
believe, a reasonable proxy for the correlation of orders submitted by different asset managers.\footnote{This value is indeed compatible with the
value of $\corr_{\epsilon}$ obtained with the version of the database used by~\cite{zarinelli2015beyond}, which identified metaorders
coming from the same investor.}

\subsection{Market Impact with Correlated Metaorders}
\label{signcorr}

A natural model would be to consider the $\phi_i$'s as exchangeable multivariate Gaussian variables of zero mean, variance $\volfl_N^2$ and cross-correlation coefficient $\mathbb{C}_{\phi}(N)$. 
Appendix~\ref{AppD} shows that the qualitative behavior for independent metaorders remains the same when $\mathbb{C}_{\phi}(N) > 0$. Specifically, one
finds that the average impact $I_N(\phi)$ can be obtained 
by making the substitution 
\begin{equation}
\phi \to \phi [1 + (N-1) \mathbb{C}_{\phi}(N)].
\end{equation}
in the expression of $I_N(\phi)$ for independent Gaussians. This is expected, as $(N-1)\mathbb{C}_{\phi}(N)$ gives the effective number of additional
volume-weighted metaorders correlated to the original one. By the same token though, 
$I_N(\phi)$ still vanishes linearly for small $\phi$, whereas empirical data suggests a positive intercept when $\phi \to 0$.

As an alternative model that emphasizes sign-correlations, let us assume that the joint distribution of the $\phi_i$'s can be written as
\begin{equation}
P(\bphi_N)=\mathcal{P}(\bm{\epsilon}_N) \prod_{i=1}^N p(|\phi_i|),
\end{equation}
meaning that metaorder sizes are independent, while the signs are possibly correlated. This specific form is motivated by the observation that the
size of a metaorder is mainly related
to the assets under management of the corresponding financial institution, while the sign is related to the trading signal. One can expect that different investors use correlated information sources, 
while the size of the trades is idiosyncratic.

\begin{figure}
\begin{center}
\includegraphics[width=1.0\textwidth]{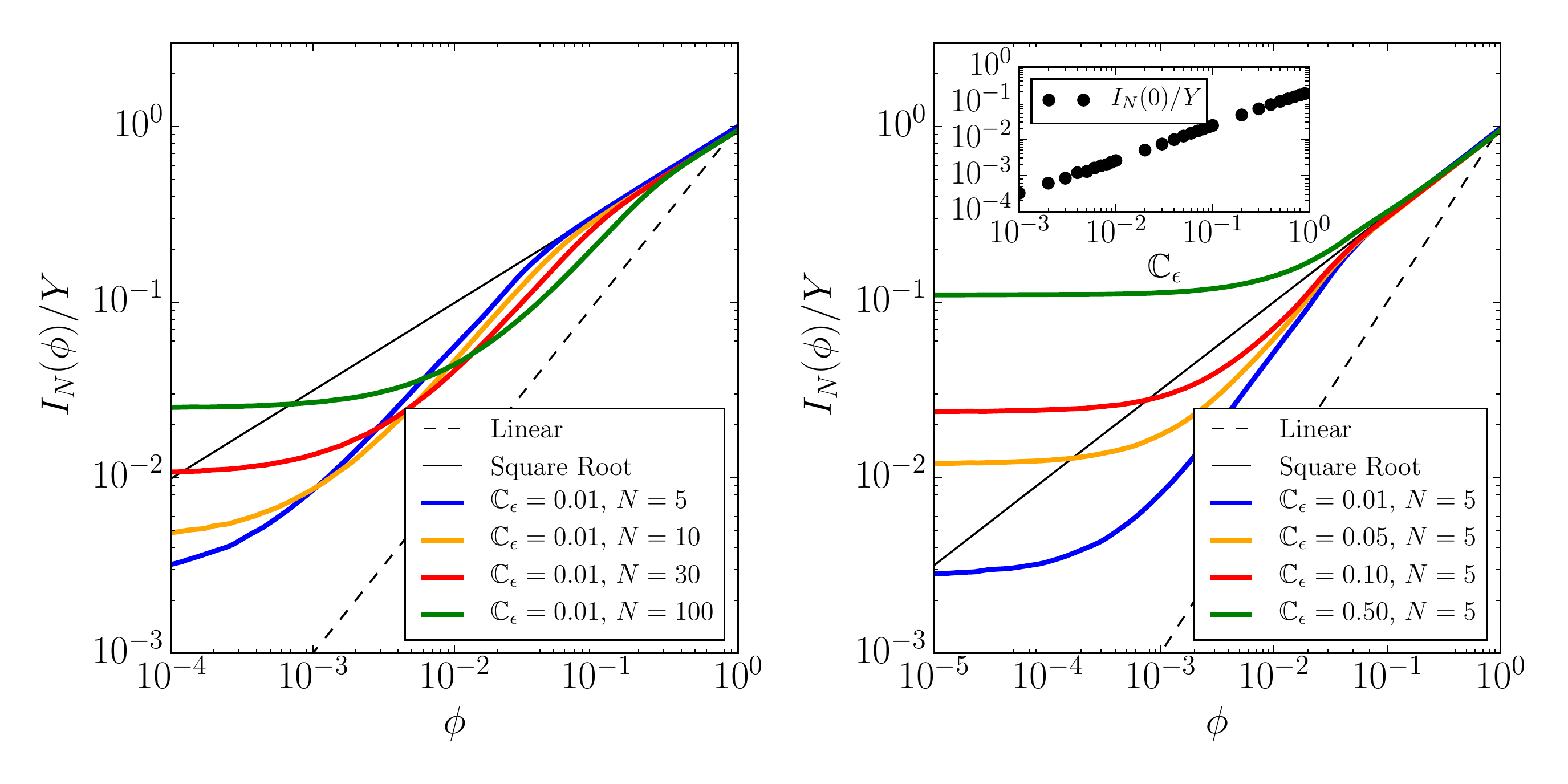}
\caption{Market impact curves $I_N(\phi)/Y$ computed for the correlated signs model, where volumes are drawn according to a half-normal distribution with
$\avg[|\phi_i|]=\volfl \sqrt{2/\pi}$ and $\volfl=0.8\%$ is set equal to its empirical value averaged over $N$. 
(Left panel) Numerical simulations for fixed sign correlation  $\corr_{\epsilon}=0.01$ but varying number of metaorders
$N\in\{5, 10, 30, 100\}$. (Right panel) Numerical simulations for fixed number of metaorders $N=5$ but with varying sign correlation
$\corr_{\epsilon} \in\{0.01, 0.05, 0.1, 0.5\}$: as shown in the inset the intercept $I_N(0)/Y$ decreases linearly with the sign correlation for  $\corr_{\epsilon}\rightarrow 0$. Note that the individual price impact $I_N(\phi)$
values converges to a positive constant when $\phi \to 0$. For intermediate $\phi$, $I_N(\phi)$ is linear and crosses-over at larger $\phi$ to a square root.}
\label{figure11}
\end{center}
\end{figure}

We further assume that there is a unique common factor determining the sign of the metaorders. In other words, the statistical model for the signs is the following:
\begin{equation}
\mathbb{P}(\epsilon_i=+1|\tilde{\epsilon})= \frac{1}{2} (1 +{\gamma}_\epsilon \tilde \epsilon); \qquad \mathbb{P}(\epsilon_i=-1|\tilde{\epsilon})= \frac{1}{2} (1 - {\gamma}_\epsilon \tilde \epsilon),
\end{equation}
where $\tilde \epsilon$ is the hidden sign factor, such that $\mathbb{P}(\tilde \epsilon= \pm 1)=1/2$, and ${\gamma}_{\epsilon}$ is the sign correlation between each sign $\epsilon_i$ and the hidden sign factor $\tilde{\epsilon}$. A simple calculation leads to
\begin{equation}
\corr_{\epsilon}(N)=\mathbb{P}(\epsilon_i=\epsilon_j)-\mathbb{P}(\epsilon_i=-\epsilon_j)={\gamma}^2_\epsilon.
\label{corrsign}
\end{equation}
where we omitted the $\gamma_\epsilon$'s explicit dependence on $N$.
Contrarily to the Gaussian case, we have not been able to obtain analytical formulas, but instead relied on numerical simulations to
obtain $I_N(\phi)$ for different combinations of $\corr_\epsilon(N)$ and $N$, reported in Fig.~\ref{figure11}. Results for unsigned volumes generated from a half-normal distribution calibrated on data are
shown in Fig.~\ref{figure11}. We observe that the individual price impact $I_N(\phi)$ converges to a positive constant $I_N(0) > 0$ when $\phi \to
0$, despite $I_N(0)=0$ for a Gaussian model. 
For intermediate $\phi$, $I_N(\phi)$ is linear and it crosses over at larger $\phi$ to a square root. For fixed $N$ the intercept value increases with
the sign correlation $\corr_{\epsilon}$, see the right panel of Fig.~\ref{figure11}. The intuition is that conditioned to the fact that I buy, and
independently of the size of my trade,
the order flow of other actors will be biased towards buy as well, and I will suffer from the impact of their trades. In fact, subtracting the
non-zero intercept of $I_N(\phi)$ leads to impact curves that look almost identical to those of Fig.~\ref{fig:nonintera}, i.e. a linear region for
small $\phi$ followed by a square root region beyond a crossover value
$\phi_N^* \sim  \Sigma_N \, \sqrt{N-1}$. Since for large $N$ $\phi_N^* \to 0$, one simply expects a square-root law, shifted by the intercept $I_N(0)$.

\subsection{Empirical Calibration of the Model}
\label{sec:confronting}

\begin{figure}
\begin{center}
\includegraphics[width=0.5\linewidth]{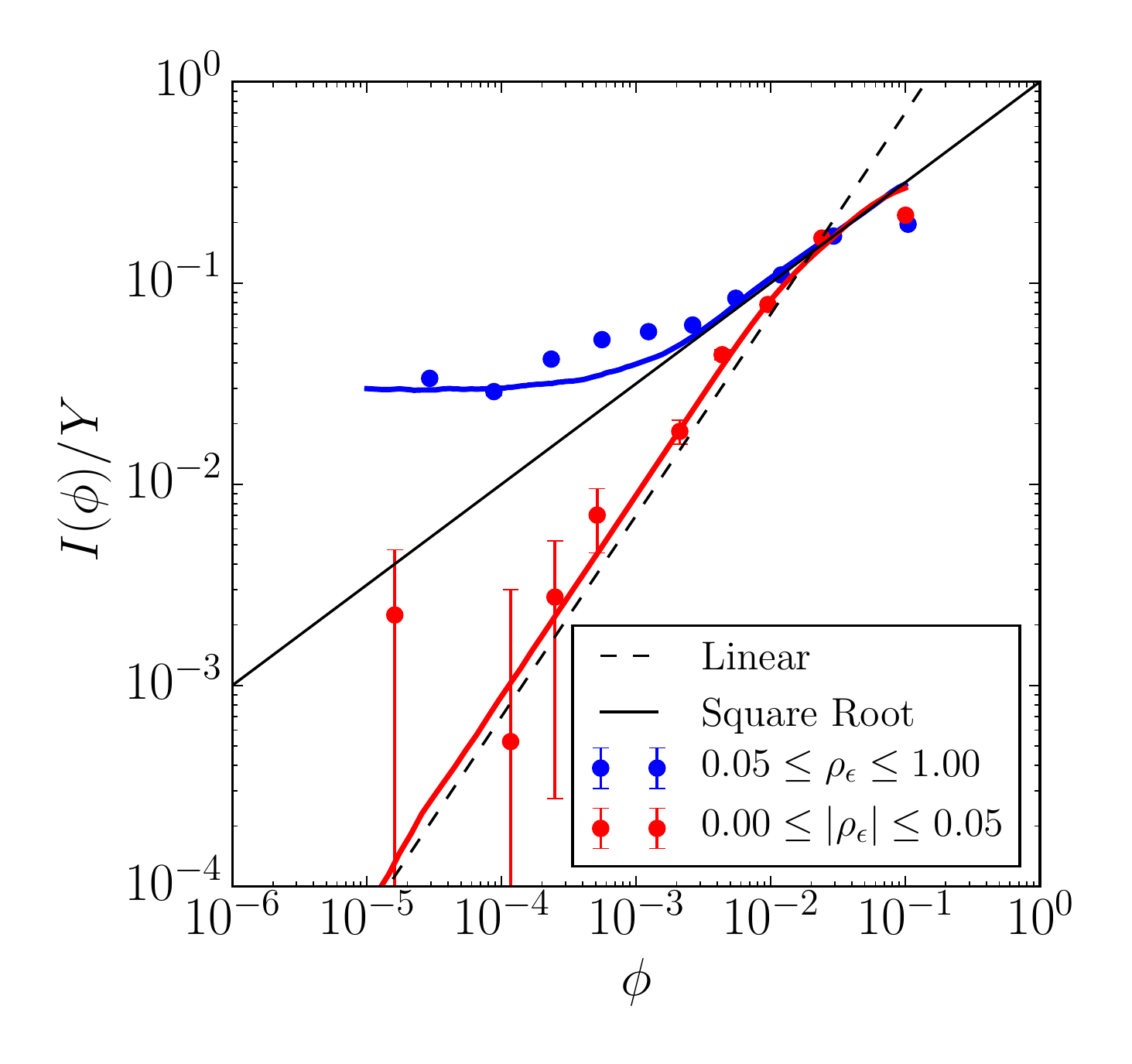}
\caption{Comparison between calibrated sign-correlated model (colored lines) and empirical data (circles) for the case $N\in[2,10]$: The sample is split into two sub-samples respectively with 
$\rho_{\epsilon} \ge 0.05$ and $0\le |\rho_{\epsilon}|\le 0.05$. The theoretical curves are calculated through numerical simulations as explained in the main text (see Appendix~\ref{app:calibsign} for all the details of the procedure).}
\label{fig:thvsemp}
\end{center}
\end{figure}

With the aim to compare the model prediction with empirical data, we propose a calibration method described in Appendix ~\ref{app:calibsign}. This is based on
the assumption that metaorder signs are independent random variables sampled from a half-Gaussian calibrated on empirical data. The metaorder sign
correlation structure can be estimated by introducing a realized sign correlation
\begin{equation}
  \label{eq:rhosign}
  \rho_\epsilon := \frac{2}{N(N-1)} \sum_{1\leq i<j \leq N} \epsilon_i \epsilon_j,
\end{equation}
which is then used to estimate the sign correlation $\corr_\epsilon (N)$ of Eq.~\eqref{corrsign}. Once the model is calibrated, we use numerical
simulations to compute the expected market impact $I(\phi)$, see Appendix~\ref{app:calibsign} for the precise details of the procedure. 

Fig.~\ref{fig:thvsemp} shows that imposing correlation only between the signs leads to a very good prediction of the empirical curves, justifying the adoption of the sign correlated model. All the features of the empirical impact curves are qualitatively well reproduced, in line with Fig.~\ref{fig:IxpPXCO}. This includes the clear deviations from the square root law for $\phi \le 10^{-3}$ with both a linear regime and a constant price impact $I_0$ when $\phi \to 0$.

\section{Conclusions}
\label{sec:conclusions}
It is a commonly acknowledged fact that market prices move during the execution of a trade -- they increase (on average)
for a buy order and decrease (on average) for a sell order. This is, loosely stated, the phenomenon known as market impact.  
In this paper we have presented one of the first studies breaking down market impact of metaorders executed  by different investors, and taking into account 
interaction/correlation effects. We investigated how to aggregate the impact of individual actors in order to best explain the daily price moves. The large number
and heterogeneity of the metaorders traded by financial institutions allows precise measurements of price
impact in different conditions with reduced uncertainty. We found that both the number of
actors simultaneously trading on a stock and the {\it crowdedness} of their trade (measured by the correlation of metaorder signs) 
are important factors determining the impact of a given metaorder.  

Our main conclusions are as follows:
\begin{itemize}
\item The market chiefly reacts to the total net order flow of ongoing metaorders, the functional
form being well approximated by a square root at least in a range of volume fraction $\phi$. As expected in anonymous markets, impact is insensitive
to the way order flow is distributed across different investors.
\item The number $N$ of executed metaorders and their mutual sign correlations $\corr_\epsilon$ are relevant parameters
when an investor wants to precisely estimate the market impact of their own metaorders. 
\item Using a simple heuristic model calibrated on data, we are able to reproduce to a good level of precision 
the different regimes of the empirical market impact curves, as a function of $\phi$, $N$ and $\corr_\epsilon$. 
\item When the number of metaorders is not large, and when $\corr_\epsilon > 0$, a small investor will 
observe {\it linear} impact with a non-zero intercept $I_0$, crossing over to a square-root law at larger $\phi$. 
$I_0$ grows with $\corr_\epsilon$ and can be interpreted as the average impact of all other metaorders. 
\item When the number of metaorders is very large and the investor has no correlation with their average sign, they should 
expect on a given day a square-root impact randomly shifted upwards or downwards by $I_0$. Averaged over all days, a pure square-root law 
emerges, which explains why such behavior has been reported in many empirical papers. 
\end{itemize}
On the last point, we believe that our study sheds light on an apparent paradox: How can a non-linear impact law survive in the presence of a large
number of simultaneously executed metaorders?  As we have seen, the reason is that for a metaorder uncorrelated with the rest of the market, the
impacts of other metaorders cancel out on average. Conversely, any intercept of the impact law can be
interpreted as a non-zero correlation with the rest of the market.

Given the importance of the subject, our results present several interesting applications. Our aggregated price impact model should be of interest both to practitioners 
trying to monitor and reduce their trading costs, and also to regulators that seek to improve the stability of markets.

\section*{Acknowledgments}
The authors thank Michael Benzaquen (who participated to the early stage of this project), Stanislao Gualdi, P\'eter Horvai, Felix Patzelt, Bence T\'oth and Elia Zarinelli for critical discussions of the topic.

\appendix
\section{Metaorder Statistics}
\label{app:metastat}

\begin{figure}
\begin{center}
\includegraphics[width=0.85\textwidth]{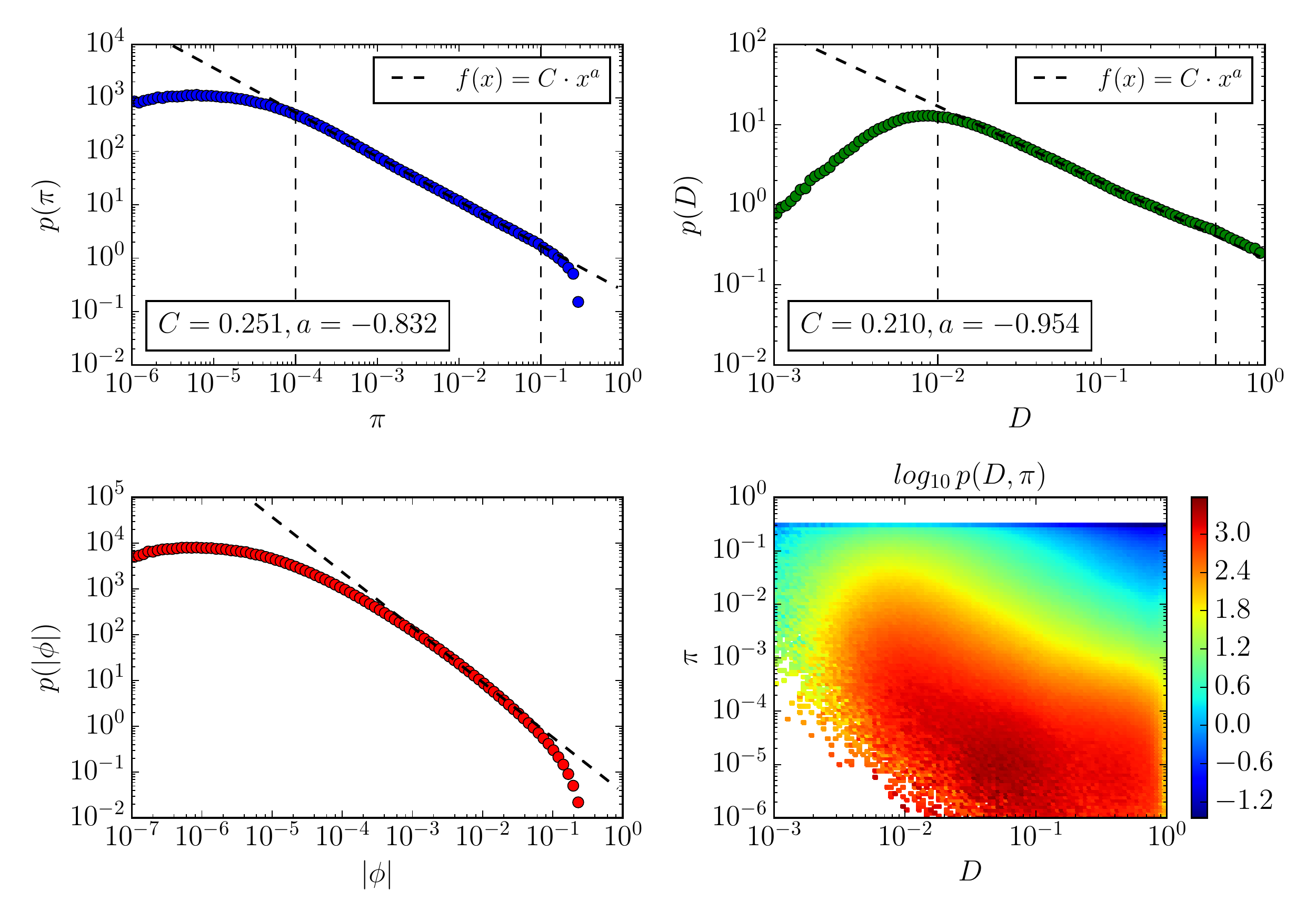}
\caption{Estimation of the probability density function of the participation
rate $\pi$  (top left),
duration $D$ (top right) and unsigned daily fraction $|\phi|$ (bottom left). All these
panels are in log-log scale.
The first two also show the best fit with a power-law function in the region
bounded by the vertical dashed lines. The bottom right panel shows the logarithm
of the estimated joint probability density function $p(D,\pi)$ in double
logarithmic scale of the duration $D$ and of the participation rate $\pi$. These statistics are in agreement with the ones
obtained in Ref.~\cite{zarinelli2015beyond}.}
\label{dist1}
\end{center}
\end{figure}

We now describe some statistics of the metaorders. The participation rate $\pi$
is defined as the ratio between the number of shares $|Q|$ traded by the metaorder and the
whole market during the execution interval $[t_{\mathrm s}, t_{\mathrm e}]$
\begin{equation}
\pi=\frac{|Q|}{V(t_{\mathrm e})-V(t_{\mathrm s})}.
\end{equation}
The duration $D$ expressed in volume time is equal to
\begin{equation}
D=\frac{V(t_e)-V(t_s)}{V(t_c)},
\end{equation}
while the unsigned daily fraction is defined as the ratio between the metaorder unsigned volume $|Q|$ and the volume
$V(t_c)$ traded by the market in the whole day:
\begin{equation}
 |\phi|=|Q|/V(t_c)=\pi \cdot D.
\end{equation}

We find that the participation rate $\pi$ and the duration $D$ are both well approximated by
truncated power-law distributions over several orders of magnitude. The
estimated probability density function of the participation rate $\pi$ is shown
in log-log scale in the top left panel of Fig.~\ref{dist1}. A power-law fit in
the region $10^{-4}\le \pi \le 10^{-1}$, i.e. over three orders of magnitude
gives a best fit exponent $a=-0.832 \pm 0.001$. The top right panel of
Fig.~\ref{dist1} shows the estimated probability density function of the
duration $D$ of a metaorder. A power-law fit in the region bounded by
vertical dashed lines ($0.01 \le D \le 0.5$) gives a power-law exponent $a=-0.954
\pm 0.002$. These power laws are very heavy-tailed, meaning that
there is substantial variability in both the participation rate and the duration.
Note that in both cases the variability is
intrinsically bounded, and therefore the power law is automatically truncated:
by definition $\pi \le 1$ and $D \le 1$. In addition, for
$p(D)$ there is a small bump on the right extreme of the distribution
corresponding to all-day metaorders. The deviation from a power law for small
$D$ is a consequence of our Filter 3 retaining only metaorders lasting at least 2
min, which in volume time corresponds to $2/390\simeq 0.005$ on  average. The
bottom left panel shows the probability density function of the unsigned daily fraction
$|\phi|$. In this case the distribution is less fat-tailed, and clearly not a power law. This is potentially an important result, as the
predictions of some theories for market impact depend on this,
and have generally assumed power-law behavior. In conclusion, the bottom right
panel of Fig.~\ref{dist1} shows the logarithm of the estimated joint
probability density function $p(D,\pi)$ in double logarithmic scale. We measure a
very low linear correlation (-0.08) between the two variables, the main
contribution coming from the extreme regions, i.e. very large $\pi$ implies very small $D$
and vice versa. In other words, as expected, very aggressive metaorders are
typically short and long metaorders more often have a small participation rate.

\section{From the Bare  to the Market Impact Function}
The expected individual market impact of a metaorder with signed volume $\phi$ is estimated by
\begin{equation}
I(\phi)= \sum_N p(N) I_N(\phi)
\end{equation} 
where $p(N)$ is the probability distribution of the daily number $N$  of metaorders per asset  and
\begin{equation}
I_N(\phi)=\avg[\imp(\bphi_N)|\phi_k=\phi]=\int \dd \phi_1 \cdots \dd \phi_N P(\bphi_N|\phi_k=\phi) \left( \phi_k +\sum_{i\neq k}^N \phi_i \right)^{\bullet 1/2}
\label{Inphi}
\end{equation}
is the market impact  computed from the bare impact function $\imp(\bphi_N):=(\sum_{i=1}^N \phi_i)^{\bullet 1/2}$ with  fixed $N$. Assuming that $p(N)$ is kwown  a priori  the market impact $I_N(\phi)$ is given from the expectation in Eq. \eqref{Inphi} done over the conditional probability distribution $P(\bphi_N|\phi_k=\phi)$ of the volume metaorders.
However,  in  such a way to do  analytical computation it is necessary to assume reasonable hypothesis for  the  joint distribution function $P(\bphi_N):=P(\phi_1,\cdots,\phi_N)$.

For this reason, we start considering the case of i.i.d. metaorders and we firstly show analytically  how the transition from the square root to a linear market impact is possible in the i.i.d. Gaussian framework. Secondly,  we generalize these results in the limit of large $N$  for any symmetric volume distribution which sastifies  the Central Limit Theorem assumptions. Thirdly, we show that the same results continue to be valid introducing correlation between the signed volumes in a Gaussian framework. For last, we describe how to compute numerically the market impact in Eq. \eqref{Inphi} in the case of metaorder signs correlated and i.i.d. unsigned volumes.

\subsection{Market Impact with IID Metaorders}
\label{app:calc}
In the case of  i.i.d.  signed metaorders, the conditional joint distribution  factorizes as  
\begin{equation}
P(\bphi_N|\phi_k=\phi)=\prod_{i \neq k}^Np(\phi_i);
\end{equation}
this implies that  the price impact $I_N(\phi)$ of   a single metaorder $\phi_k=\phi$ out of $N$ is given by
\begin{equation}
I_N(\phi)=\int   \dd\phi_m \underbrace{\frac{1}{2\pi} \int \dd\lambda  e^{-i\lambda \phi_m} \hat{p}(\lambda)^{N-1}}_{p(\phi_m)} (\phi+\phi_m)^{\bullet 1/2}=\int \dd  \phi_m p(\phi_m) (\phi+\phi_m)^{\bullet 1/2}
\label{Eq:convol}
\end{equation}
where $\phi_m= \sum_{i \neq k}^N \phi_i$ is the net order flow executed simultaneously to  the metaorder $\phi_k$ and $\hat{p}(\lambda)=\avg[e^{i\lambda \phi_i}]$ is the characteristic function of the signed volume distribution $p(\phi_i)$. Although the introduction of the characteristic function $\hat{p}(\lambda)$ in Eq. (\ref{Eq:convol}) will be a convenient way to exploit the convergence of the net order flow distribution $p(\phi_m)$ as discussed in Appendix \ref{app:calciidnongauss}, the computation of the market impact $I_N(\phi)$  in a analytical closed form is possible only in the Gaussian case as shown in the next Appendix \ref{app:calciidgauss}.

\subsubsection{Independent Gaussian Metaorders}
\label{app:calciidgauss} 
In the  gaussian case, i.e. $p(\phi_i)\sim\mathcal{N}(0,\volfl_N^2)$ with $\volfl_N^2=\var[\phi_i|N]$, we can go further analytically in Eq. \eqref{Eq:convol} since $\hat{p}(\lambda)=e^{-\volfl_N^2\lambda^2/2}$. In fact the integral representation of the  price impact  
\begin{multline}
I_N(\phi)=
\frac{1}{\volfl_N \sqrt{2\pi  (N-1)}}\int_{-\infty}^{\infty}   \dd\phi_m
e^{-\phi_m^2/(2\volfl_N^2(N-1))} (\phi+\phi_m)^{\bullet 1/2} =\\
=\frac{1}{ \volfl_N \sqrt{2\pi (N-1)}}\int_{0}^{\infty}   \dd x \sqrt{x}
\Big[e^{-(x-\phi)^2/(2\volfl_N^2(N-1))}-e^{-(x+\phi)^2/(2\volfl_N^2(N-1))}\Big]
\label{Eq:2}
\end{multline}    
can be expressed in the following analytical way 
\begin{equation}
I_N(\phi)= \frac{\Gamma(1/4)}{2 \sqrt{\pi}} \frac{\phi}{(2(N-1)\Sigma_N^2)^{1/4}} \hspace{0.1cm}e^{-\frac{\phi^2}{2(N-1)\Sigma_N^2}}  \prescript{}{1}{F}^{}_{1}\left(\frac{5}{4},\frac{3}{2},\frac{\phi^2}{2(N-1)\Sigma_N^2}\right)
\label{Eq:2a}
\end{equation}
where $\Gamma(z)=\int_{0}^{\infty} x^{z-1}e ^{-x} \dd x$ is the Gamma function and  
\begin{equation}
_1F_1\left(\frac{5}{4},\frac{3}{2},z \right)=\frac{\Gamma(\frac{3}{2})}{\Gamma(\frac{5}{4})}\sum_{j=0}^{\infty} \frac{\Gamma(\frac{5}{4}+j)}{\Gamma(\frac{3}{2}+j)}\frac{z^j}{j!}
\end{equation}   
 is the Kummer  confluent hypergeometric function  with $z=\frac{\phi^2}{(2(N-1)\Sigma_N^2)}$\cite{wolfram1991mathematica}.

\begin{figure}
\begin{center}
\includegraphics[width=0.5\linewidth]{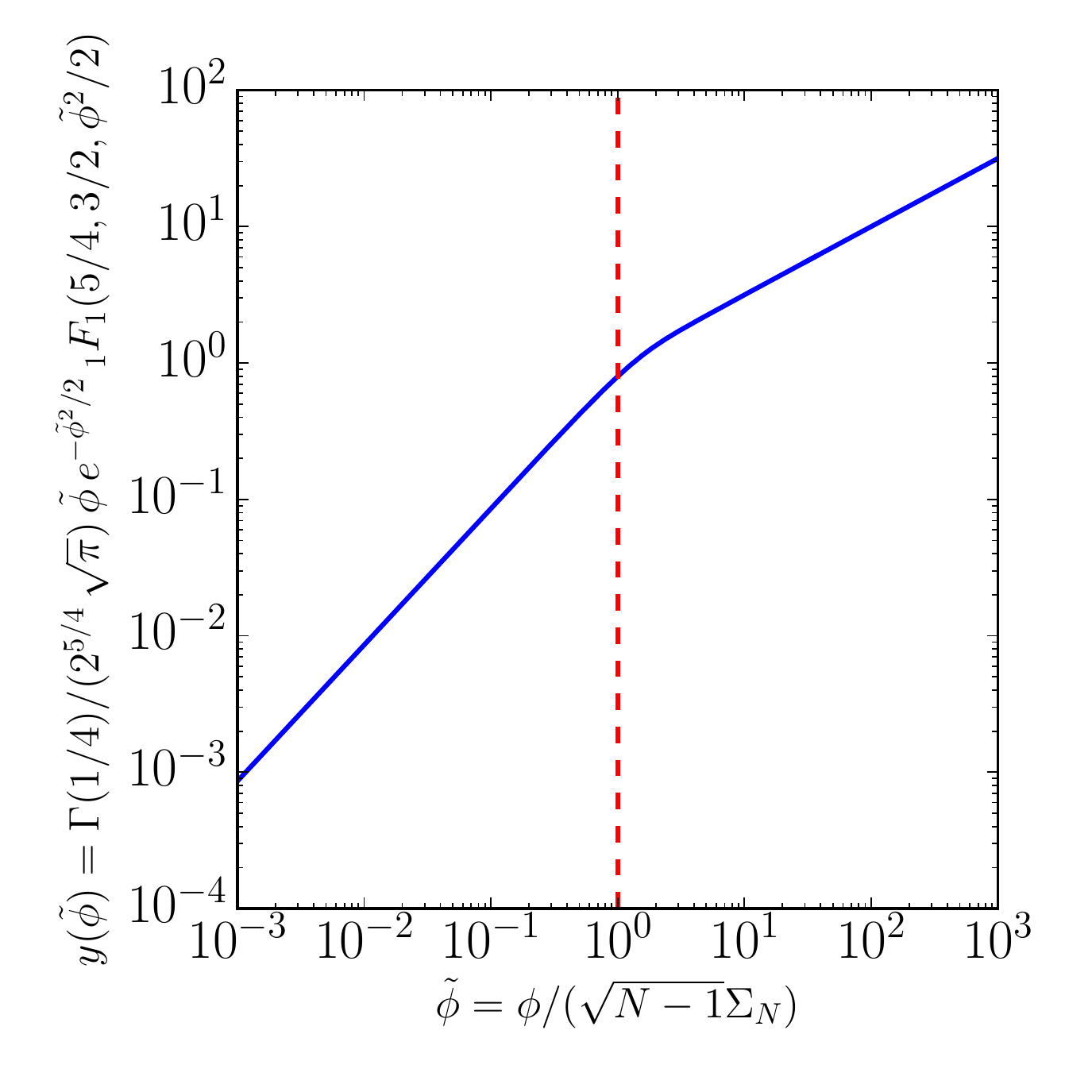}
\caption{Rescaled market impact function  $y(\tilde{\phi})=I_N(\phi)/((N-1)\Sigma_N^2)^{1/4}$ in function of  the dimensionless parameter $\tilde{\phi}:=\phi/(\sqrt{N-1}\Sigma_N)$ with $I_N(\phi)$  given by Eq. \eqref{Eq:2a}: the vertical dashed red line represents the transition from the linear impact (left side) to the square-root one  (right side).}
\label{fig:Appendix}
\end{center}
\end{figure}

The  price impact $I_N(\phi)$ in Eq. \eqref{Eq:2a} is shown in the left panel of Fig. \ref{fig:nonintera} for different $N$ and the parameter $\volfl_N$ \textit{fixed}. If the metaorder volume $\phi$ is smaller than the sum of the other $N-1$ metaorders, i.e. $\phi \ll \phi_m$, then price impact is linear. Instead, when our
metaorder dominates, i.e. $\phi \gg \phi_m$, the price impact follows a square root function.
The transition from the
linear to the square root regime takes place around $\phi_N^* \simeq \volfl_N
\sqrt{N-1}$, where $\volfl_N \sqrt{N-1}$ is naturally interpreted as a measure for the market noise, in agreement with the change of the functional shape of the rescaled market impact function $y(\tilde{\phi})=I_N(\phi)/((N-1)\Sigma_N^2)^{1/4}$ represented in Fig. \ref{fig:Appendix}  in function of the adimensional parameter $\tilde{\phi}:=\phi/(\sqrt{N-1} \Sigma_N)$. To note furthermore that  the linear regime comes out immediately from the  expansion of $I_N(\phi)$  in  Eq. \eqref{Eq:2a} around $\phi=0$ as follows 
\begin{multline}
\small
\hspace{-1.5cm}
I_N(\phi)= \frac{1}{\volfl_N \sqrt{2\pi  (N-1)}}  \int_0^{\infty} \dd x \sqrt{x}
e^{-x^2/(2\volfl_N^2(N-1))}\Big[e^{-(\phi^2-2x\phi)/(2\volfl_N^2(N-1))}-e^{-(\phi^2+2x\phi)/(2\volfl_N^2(N-1))}\Big] \simeq \\
\simeq \sqrt{\frac{2}{\pi}}\frac{ \phi}{ (\volfl_N^2(N-1))^{3/2}} \int_{0}^{\infty} \dd x x^{3/2} e^{-x^2/(2\Sigma_N^2 (N-1))} =2^{3/4}\frac{\Gamma(5/4)}{(\pi^2
\Sigma_N^2 (N-1))^{1/4}}\phi.
\label{expansionphi0}
\end{multline}

\begin{definition}
\label{r1}
It follows that for i.i.d. Gaussian metaorders the slope of the linear price impact  region decreases with $N$ (as shown explicitly in Eq. \eqref{expansionphi0}) and  the crossover to the square root region happens in  $\phi_N^*$ obtained by solving
\end{definition}

\begin{equation}
\xi \frac{\phi_N^* }{(\volfl_N^{2}(N-1))^{1/4}} \simeq
(\phi_N^*)^{1/2},
\label{xi}
\end{equation}
i.e. $\phi_N^* \simeq \xi^{-1} \volfl_N \sqrt{N-1}$ with  $\xi=2^{3/4} \Gamma(5/4)/\sqrt{\pi}$.

\subsubsection{Limit of Large $N$  for Generally Distributed Independent Metaorders}
\label{app:calciidnongauss}
The previous conclusions discussed in the i.i.d.  Gaussian framework are also valid for others i.i.d. volume distributions as discussed in this Appendix: in fact, we can generalize them in the limit of large $N$ for any symmetric volume distribution $p(\phi_i)$ which sastifies  the Central Limit Theorem assumptions.

In the limit of large $N$, for which the Central Limit Theorem applied if certain conditions (discussed below) on $p(\phi_i)$ are sastified,  the symmetric volume distribution $p(\phi_m)$ introduced in Eq. \eqref{Eq:convol} converges  to a stable law  $\mathcal{G}_\alpha$ described by a characteristic function 
\begin{equation}
\hat{p}(\lambda)=\avg[e^{i\lambda \phi_m}]=e^{-c |\lambda|^{\alpha}}
\label{Eq:cf}
\end{equation}
where $c \in (0,\infty)$  is the scale parameter and $\alpha \in (0,2]$  is the
stability exponent. In other words we say that the volume distribution $p(\phi_i)$ belongs to the \textit{domain of attraction} of the stable distribution $\mathcal{G}_{\alpha}$ if there exist constants $a_m \in \mathbb{R}$, $b_m>0$ such that 
\begin{equation}
b_m^{-1}(\phi_m - a_m) \rightarrow \mathcal{G}_{\alpha},
\end{equation}
i.e. the renormalized and recentred sum $\phi_m= \sum_{i \neq k}^N \phi_i$  converges  in distribution to $\mathcal{G}_{\alpha}$.
The Central Limit Theorem gives the conditions such that this convergence in distribution to a stable law $\mathcal{G}_\alpha$ is guaranted:
\begin{itemize}
\item  $p(\phi_m)$ converges to  a Gaussian distribution ($\alpha=2$ in Eq. \eqref{Eq:cf}) if and only if 
\begin{equation}
\int_{|\phi_i|\le x} \phi_i^2 p(\phi_i) \dd \phi_i
\label{cond1}
\end{equation}
is a slowly varying function $L(x)$, i.e. $\text{lim}_{x \rightarrow \infty} L(t x)/L(x)=1$ for all $t >0$;  then it follows that if $\Sigma_N^2=\var[\phi_i|N]<\infty$
\begin{equation}
((N-1)^{1/2}\Sigma_N)^{-1}\phi_m \longrightarrow \mathcal{N}(0,1),
\end{equation}
while if $\var[\phi_i|N]=\infty$
\begin{equation}
((N-1)^{1/2}L_1)^{-1}\phi_m \longrightarrow \mathcal{N}(0,1)
\end{equation}
with $L_1$ a slowly varying function and $\mathcal{N}(0,1)$ a gaussian distribution with mean zero and variance 1. 
\item $p(\phi_m)$ converges to a L\'evy distribution  (for  some $\alpha <2$ in Eq. \eqref{Eq:cf})  if and only if 
\begin{equation}
\int_{-\infty}^{-x} p(\phi_i) \dd \phi_i=\frac{c_1+o(1)}{x^{\alpha}} L(x), \quad 1-\int_{-\infty}^{x}p(\phi_i) \dd \phi_i=\frac{c_2+o(1)}{x^{\alpha}}L(x), \quad x \rightarrow \infty
\label{cond2}
\end{equation}
where $L(x)$ is a slowly varying function and $c_1$, $c_2$ are non-negative constants such that $c_1+c_2>0$; then  it follows that 
\begin{equation}
((N-1)^{1/\alpha}L_2)^{-1}\phi_m \longrightarrow \mathcal{L}(c,\alpha)
\end{equation}
with $L_2$ a slowly varying function and $\mathcal{L}(c,\alpha)$ a L\'evy distribution with scale parameter $c\in (0,\infty)$ and stability exponent  $\alpha \in (0,2)$.
\end{itemize}

\begin{definition}
It follows immediately that for  any volume distributions $p(\phi_i)$  belonging to the \textit{domain of attraction} of the normal law, i.e. satisfying the condition in Eq. \eqref{cond1}, the price impact $I_N(\phi)$  in the limit of large $N$  is described by Eq. \eqref{Eq:2a}  and the transition from a linear to a square root price impact discussed in the Remark \ref{r1} continue to be valid: to note that in the  case of $\var[\phi_i|N]=\infty$ it is sufficient to substitute $\Sigma_N$ in Eq. \eqref{xi} with the appropriate slowly varying function  $L_1$. 
\end{definition}

Moreover we can  show that the transition from a linear  to  a square root regime is still present  for  volume distributions $p(\phi_i)$ belonging to the \textit{domain of attraction} of the L\'evy distribution, i.e. that sastify the condition in Eq. \eqref{cond2} with $\alpha <2$ and then $p(\phi_m) \rightarrow \mathcal{L}(c (N-1)^{1/\alpha},\alpha)$. In fact, expanding to first order the bare impact function $\imp(\phi,\phi_m)=(\phi+\phi_m)^{\bullet 1/2}$ around $\phi=0$ in Eq. \eqref{Eq:convol}
\begin{eqnarray}
I_N(\phi)\simeq \frac{\phi}{2} \int 
 \dd\phi_m p(\phi_m)  
|\phi_m|^{-1/2}=\frac{\phi}{4\pi}  \int \dd \lambda \hat{p}(\lambda) \int \dd \phi_m    e^{-i\lambda \phi_m} |\phi_m|^{-1/2},
\label{deltaqo}
\end{eqnarray}
introducing the characteristic function of the net order flow $\phi_m$ given by
\begin{equation}
\hat{p}(\lambda)=e^{-c (N-1)^{1/\alpha}|\lambda|^{\alpha}}
\end{equation}
for large $N$  and taking present  that
 the last integral in Eq. \eqref{deltaqo} is a known Fourier transform 
\begin{equation}
\int_{-\infty}^{+\infty} \dd\phi_m e^{-i\lambda \phi_m}
|\phi_m|^{-1/2}=\frac{\sqrt{2\pi}}{|\lambda|^{1/2}}
\end{equation}
we obtain a linear price impact
\begin{equation}
I_N(\phi)\simeq \frac{\phi}{2\sqrt{2\pi}}
\int_{-\infty}^{+\infty} \hat
p(\lambda) |\lambda|^{-1/2} \dd\lambda.
\label{eqintermedia}
\end{equation}
Though it is not possible to analytically compute the last integral,  its behavior for large $N$ is known  using the saddle point
approximation or Perron's method: since  all the
conditions of the theorem at page 105 of Ref. \cite{book1} are satisfied, one can approximate the integral in Eq. \eqref{eqintermedia} as follows
\begin{equation}
\int_{-\infty}^{+\infty} \hat p(\lambda) |\lambda|^{-1/2} \dd\lambda \sim
\Gamma\left(\frac{1}{2\alpha}\right)\frac{1}{\alpha
c^{1/(2\alpha)}(N-1)^{1/(2 \alpha)}}.
\end{equation}

\begin{definition}
In the limit of large $N$ it is then  possible to  show analytically that  for  volume distributions $p(\phi_i)$ belonging to the \textit{domain of attraction} of the L\'evy distribution the price impact is characterized by a linear regime described by  
\begin{equation}
I_N(\phi)\simeq \frac{1}{2\sqrt{2\pi}} \Gamma\left(\frac{1}{2\alpha}\right)\frac{\phi}{\alpha [c(N-1)]^{1/(2\alpha)}},
\end{equation}
 followed by a transition to a square root one around $\phi_N^*\simeq (c(N-1))^{1/\alpha}$.
\end{definition}

\subsection{Market Impact with Correlated Gaussian Metaorders}

\label{AppD}
With the aim to introduce correlations between the metaorder volumes
$\bphi_N=(\phi_1,\cdots,\phi_N)$ in  the Gaussian framework it is useful to  define the following joint probability distribution 
\begin{equation}
\small{
P(\bphi_N)=\frac{1}{Z_N}\exp\left(-\frac{A_N}{2}\sum_{i=1}^{N}\phi_i^2+\frac{B_N}
{N}\sum_{i<j}^N\phi_i\phi_j+\mu\sum_{i=1}^N\phi_i\right),
}
\label{distprob}
\end{equation}
where $Z_N$ is a normalization function, $A_N$ and $B_N$ are 
parameters depending on $N$ and $\mu$ is an external field.

\subsubsection{Calibration from Data: Means and Correlations}

The first step to calibrate $P(\bphi_N)$ in Eq. \eqref{distprob} is to express the model parameters $A_N, B_N$ and $\mu$ in terms of observable quantities, namely
$\mathbb{E}[\phi_i\phi_j|N]$ and $\mathbb{E}[\phi_i|N]$.
Due to the presence of an interaction term, the computation of  $Z_N$ requires the use of a Hubbard-Stratonovich transformation (valid only for
$B_N>0$):
\begin{equation}
\exp\left(\frac{B_N}{2N}\sum_{i,j}^N\phi_i\phi_j\right)=\int_{-\infty}^{\infty}
\frac{dy}{\sqrt{2\pi/NB_N}}\exp\left(-\frac{NB_Ny^2}{2}+B_N\sum_{i=1}^N\phi_iy\right).
\end{equation}
This allows us to rewrite the probability distribution in Eq. \eqref{distprob} as
\begin{equation}
P(\bphi_N)=\frac{1}{Z_N}\sqrt{\frac{NB_N}{2\pi}}\int dy \prod_{i=1}^N 
\exp\left[-\frac{1}{2}\left(A_N+\frac{B_N}{N}\right)\phi_i^2+(\mu+B_Ny)\phi_i-\frac{NB_N}
{2}y^2\right] \, .
\label{PQN}
\end{equation}
The partition function then reads
\begin{equation}
Z_N=\left[\frac{2\pi}{\left(A_N+\frac{B_N}{N}\right)}\right]^{N/2}\sqrt{\frac{A_NN+B_N}{
A_NN+B_N(N-1)}}\exp\left[\frac{N^2\mu^2}{2(A_NN+B_N(N-1))}\right]
\label{Eq:Z}
\end{equation}
valid only for $B_N<A_N+B_N/N$. Eq.~\eqref{Eq:Z} can be used  to derive the following relations:
\begin{eqnarray}
\frac{\partial \log Z_N}{\partial \mu} &=& N \mathbb{E}[\phi_i|N], \\
\frac{\partial \log Z_N}{\partial A_N} &=& -\frac{N}{2}\mathbb{E}[\phi_i^2|N], \\
\frac{\partial \log Z_N}{\partial B_N} &=& \left(\frac{N-1}{2}\right)
\mathbb{E}[\phi_i\phi_j|N] \hspace{0.5cm}\textrm{with}\hspace{0.5cm}(i\neq j)
\end{eqnarray}
which assuming symmetric volumes ($\mathbb{E}[\phi_i|N]=0$, i.e. $\mu=0$)  are equivalent to
\begin{equation}
\mathbb{E}[\phi_i^2|N]=\frac{A_N+2B_N/N-B_N}{(A_N+B_N/N-B_N)(A_N+B_N/N)}
\label{Q2_bis}
\end{equation}
and
\begin{equation}
\mathbb{E}[\phi_i\phi_j|N]=\frac{B_N/N}{(A_N+B_N/N-B_N)(A_N+B_N/N)}.
\label{QiQj_bis}
\end{equation}
Furthermore, combining Eqs.~\eqref{Q2_bis} and \eqref{QiQj_bis} we can derive  the volume correlation
\begin{equation}
\corr_{\phi}(N)=\frac{\mathbb{E}[\phi_i\phi_j|N]-\mathbb{E}[\phi_i|N]^2}{\mathbb{E}[
\phi_i^2|N]-\mathbb{E}[\phi_i|N]^2}=
\frac{\mathbb{E}[\phi_i\phi_j|N]}{\mathbb{E}[\phi_i^2|N]}=\frac{B_N/N}{A_N+2B_N/N-B_N}.
\label{correla}
\end{equation}
Vice versa,  from  Eqs.~\eqref{Q2_bis} and ~\eqref{correla} we can obtain for the model parameters
\begin{equation}
A_N=\frac{1-2\corr_{\phi}(N)+N\corr_{\phi}(N)}{(1-\corr_{\phi}(N))(1-\corr_{\phi}(N)+N \corr_{\phi}(N)) \mathbb{E}[\phi_i^2|N]}
\label{A}
\end{equation}
and
\begin{equation}
B_N=\frac{N \corr_{\phi}(N)}{(1-\corr_{\phi}(N))(1-\corr_{\phi}(N)+ N \corr_{\phi}(N))  \mathbb{E}[\phi_i^2|N]},
\label{B}
\end{equation}
which are useful to fit the Gaussian model to data. We will use Eq. \eqref{A} and Eq. \eqref{B} to estimate $A_N$ and $B_N$, replacing correlations and expectations by their empirical natural counterpart. The properties of this kind of estimators, belonging to the GMM (Generalized Method of Moments) is for instance discussed in Ref. \cite{GMM}.   

\subsection{Analytical Computation of Market Impact}
To compute analytically the market impact function $I_N(\phi)$  from Eq. \eqref{Inphi} in the Gaussian correlated framework we adopt the following strategy
\begin{enumerate}[1.]
\item firstly we factorize the joint probability distribution $P(\bphi_N)$ in Eq. \eqref{distprob} with a not-null external field $\mu \neq 0$;
\item secondly we use the trick of the previous point to compute the market impact $I_N(\phi)$ in presence of an effective field $\tilde{\mu}$ induced by the correlation of the net order flow $\phi_m=\sum_{i \neq k}^N \phi_i$ with the known metaorder of size $\phi$.
\end{enumerate}

\begin{def1}
The joint probability distribution in Eq. \eqref{distprob}  can be written in the following  matrix form 
\begin{equation}
P(\bphi_N)=\frac{1}{Z_N}\exp\left(-\frac{1}{2} \bphi_N^T \bm{\mathbb{M}}
\bphi_N +\bm{\mu}^T  \bphi_N  \right),
\label{matP}
\end{equation}
where
\begin{itemize} 
\item $\mathbb{M}$ is  a $N$x$N$ real and symmetric matrix  with  the elements  on
the principal diagonal equal to $A_N$ and the ones elsewhere equal to $-B_N/N$,
\item $\bm{\mu}$ is a N-dimensional vector with all the elements equal to the scalar $\mu \neq 0$.
\end{itemize}
Through the orthogonal transformation $\tilde{\bphi}_N=\mathbb{O}\bphi_N$ which diagonalizes the matrix $\mathbb{M}$, i.e. $\mathbb{O}^T\mathbb{M}\mathbb{O}=\mathrm{diag}(\lambda_1,\cdots,\lambda_N)$,
the joint probability distribution $P(\bphi_N)$ factorizes as  
\begin{equation}
P(\tilde{\bphi}_N)=\frac{1}{Z_N}\prod_{m=2}^N
\exp\left[-\frac{\lambda_2}{2}\tilde{\phi}_m^2\right]\exp\left[-\frac{1}{2}
\lambda_1\tilde{\phi}^2_1+\mu \sqrt{N} \tilde{\phi}_1\right]
\label{PQ}
\end{equation}
where the $N$ eigenvalues of the matrix $\mathbb{M}$ 
\begin{equation}
\lambda_1=A_N-(N-1)\frac{B_N}{N}=\frac{1}{\mathbb{E}[\phi_i^2|N] (1-\corr_{\phi}(N)+ N\corr_{\phi}(N))}
\label{h1}
\end{equation}
and
\begin{equation}
\lambda_2=\lambda_3=\dots=\lambda_N=A_N+\frac{B_N}{N}=\frac{1}{\mathbb{E}[\phi_i^2|N](1-\corr_{\phi}(N))}
\label{hm}
\end{equation}
allow us to rewrite  the partition function as
\begin{equation}
Z_N=\sqrt{\frac{2\pi}{\lambda_1}}\left[\sqrt{\frac{2\pi}{\lambda_2}}\right]^{N-1
}\exp\left[\frac{N\mu^2}{2\lambda_1}\right].
\end{equation}
In particular,  it emerges that the first component of $\tilde{\bphi}_N=\mathbb{O}\bphi_N$ is equal to
\begin{equation}
\tilde{\phi}_1=\frac{1}{\sqrt{N}}\sum_{i=1}^N \phi_i,
\end{equation}
which put in evidence that the orthogonal basis change $\bphi_N \rightarrow \tilde{\bphi}_N$ is a useful trick to compute the market impact in the context of correlated Gaussian metaorders. 
\end{def1}

\begin{def1}
To calculate the  price impact $I_N(\phi)$ defined in Eq. \eqref{Inphi} with $N$ overall correlated Gaussian metaorders and in absence of an external field
it is necessary to explicit the conditional probability distribution  
\begin{equation}
P(\bphi_N|\phi_k=\phi)=\frac{P(\phi_1,\cdots,\phi,\cdots,\phi_N)
}{p(\phi)}
\label{CPtoQ0}
\end{equation}
where $P(\phi_1,\cdots,\phi,\cdots,\phi_N)$ is given by Eq.~\eqref{distprob} setting $\mu=0$ while the marginal one is equal to
\begin{equation}
\small{
p(\phi)= \frac{1}{\sqrt{2\pi\tilde{\lambda}_1/(\lambda_1 \lambda_2)}}
\exp\left[-\frac{\phi^2}{2}\frac{\lambda_1\lambda_2}{\tilde{\lambda}_1}\right]}
,
\end{equation} 
with $\lambda_1$ and $\lambda_2$  respectively given by  Eqs.~\eqref{h1} and \eqref{hm} 
and
\begin{equation} 
\tilde{\lambda}_1=A_{N}-\frac{N-2}{N}B_{N}=\frac{1}{\mathbb{E}[\phi_i^2|N]
[1-\corr_{\phi}(N)+N\corr_{\phi}(N)][1-\corr_{\phi}(N)]} \, .
\label{ht1}
\end{equation}
It follows that  the conditional probability
distribution is equal to 
\begin{eqnarray}
\small{
P(\bphi_N|\phi_k=\phi)=\underbrace{\frac{\exp\left[-\frac{(N-1)B^2_{N}
\phi^2}{2 N^2\tilde{\lambda}_1}\right]}{(2\pi/\tilde{\lambda}_1)^{1/2}
(2\pi/\lambda_2)^{\frac{N-2}{2}}}}_{\Theta_{N}^{-1}(\phi)}} \times \\ \small{\exp\left[-\frac{A_
{N}}{2}\sum_{ i \neq k}^N\phi_i^2+\frac{B_{N}}{N}\sum_{\substack{i<j \\ i,j\neq k}}^N\phi_i\phi_j+\underbrace{\frac{\phi
B_{N}}{N}}_{\tilde{\mu}}\sum_{i\neq k}^N\phi_i\right]}
\label{PNp1Q0}
\end{eqnarray}
where it emerges that the conditioning to the metaorder with volume $\phi_k=\phi$ is equivalent to the introduction of an effective field $\tilde{\mu}$ proportional to $\phi$. 
This implies that the price impact in Eq. \eqref{Inphi} is given solving the following conditional expectation
\begin{eqnarray}
  I_N(\phi)
  &=&\mathbb{E}[\imp(\bphi_N)|\phi_k=\phi] \\
  &=&\int_{-\infty}^{\infty} \prod_{ i \neq k}^N \dd\phi_i  
\frac{P(\phi_1,\cdots,\phi,\cdots,\phi_N)}{p(\phi)} \left(\phi+\sum_{i \neq k}^N\phi_i \right)^{\bullet 1/2} \\
  &=& \frac{1}{\Theta_{N}(\phi)}\int_{-\infty}^{\infty} \prod_{ i \neq k}^N \dd\phi_i 
\exp\left[-\frac{A_{N}}{2}\sum_{i\neq k}^N\phi_i^2+\frac{B_{N}}{N}
\sum_{\substack{i<j \\ i,j \neq k}}^N\phi_i\phi_j+\tilde{\mu}\sum_{i\neq k}^N\phi_i\right] \left(\phi+\sum_{i \neq k}^N\phi_i \right)^{\bullet 1/2} \\
  &=& \frac{1}{\Theta_{N}(\phi)}\int_{-\infty}^{\infty} \prod_{ i \neq k}^N \dd\phi_i 
\exp\left[-\bphi^{*T}\mathbb{M}^*\bphi^*+\tilde{\mu}
\bphi^*\right] \left(\phi +\sum_{i \neq k}^N\phi_i \right)^{\bullet 1/2}.
\label{Eq61}
\end{eqnarray}
Herein  
\begin{itemize}
\item $\bphi^*=\{\phi_i\}_{i=1,\cdots,N}^{i \neq k}$ is a vector that contains the $N-1$ unknown metaorders volumes simultaneously executed with the one known $\phi_k=\phi$, 
\item $\mathbb{M}^*$  is a $(N-1) \times (N-1)$ symmetric and real matrix with 
$A_{N}$ on the principal diagonal and $-B_{N}/N$ elsewhere: it is easy
to check that its  eigenvalues are respectively $\lambda^*_1=\tilde{\lambda}_1$ as in Eq. \eqref{ht1} and $\lambda^*_m=\lambda_2$ as in Eq. \eqref{hm} for $m=2,\cdots,N-1$.
\end{itemize}
 As mentioned before, to solve Eq. \eqref{Eq61} it is useful to use the trick described in Step 1.: we apply in Eq.~\eqref{Eq61} the orthogonal transformation
$\tilde{\bm{\phi}}^{*}=\mathbb{H}\bphi^{*}$ that diagonalizes the matrix
$\mathbb{M}^*$  (i.e. 
$\mathbb{H}^T\mathbb{M}^*\mathbb{H}=\text{diag}(\tilde{\lambda}_1,\cdots,\tilde{\lambda}_{N-1}$)) and since the determinant of the Jacobian matrix associated to this transformation is equal to one, we obtain that
\begin{multline}
I_N(\phi)=\frac{1}{\Theta_{N}(\phi)}\int_{-\infty}^{\infty}\prod_{m=2}^Nd\tilde{\phi}^*_
m \exp\left[-\frac{\lambda_2}{2}(\tilde{\phi}^*_m)^2\right] \times \\ \hspace{1.2cm} \times \int_{-\infty}^{\infty}
d\tilde{\phi}^*_1
\exp\left[-\frac{1}{2}\tilde{\lambda}_1(\tilde{\phi}^*_1)^2+\tilde{\mu}\sqrt{N-1}\tilde
{\phi}^*_1\right](\phi+\sqrt{N-1}\tilde{\phi}^*_1)^{\bullet 1/2};
\label{deltaq1}
\end{multline}
then integrating in $\tilde{\phi}^*_m$ for $m=2,\cdots,N-1$, completing the square
in the argument of the $\exp[-\frac{1}{2}\tilde{\lambda}_1
(\tilde{\phi}^*_1)^2+\tilde{\mu}\sqrt{N-1}\tilde{\phi}^*_1] $ and doing the variable
change $x=\phi+\sqrt{N-1}\tilde{\phi}^*_1$ we derive  the following final
expression 
\begin{eqnarray}
I_N(\phi)=\frac{1}{\sqrt{2\pi (N-1)/\tilde{\lambda}_1}}\int_{0}^{\infty}\dd x \sqrt{x}
\left(\exp\left[-\frac{\tilde{\lambda}_1}{2(N-1)}\left(x-\phi(1+(N-1)
\corr_{\phi}(N))\right)^2\right]-\right. \nonumber \\
\left. \exp\left[-\frac{\tilde{\lambda}_1}{2(N-1)}
\left(x+\phi(1+(N-1)\corr_{\phi}(N))\right)^2\right]\right) \, .
\label{PIinteractive}
\end{eqnarray}
\end{def1}
\begin{definition}
From Eq. \eqref{PIinteractive} it follows that  the price impact $I_N(\phi)$ in the correlated Gaussian framework  is equivalent to the one computed
in the independent Gaussian case (see Eqs.~\eqref{Eq:2} and \eqref{Eq:2a})  substituting
\begin{equation}
\phi \longrightarrow \phi[1+(N-1) \corr_{\phi}(N)]
\end{equation}
and
\begin{equation}
\volfl_N^2 \longrightarrow 1/\tilde{\lambda}_1.
\end{equation}
\end{definition}

\subsection{Market Impact with Correlated Signs and IID\ Unsigned Volumes}
\label{app:calccorrnongauss}

In Section \ref{signcorr} we have presented a general model in which  the metaorder signs
$\epsilon_i=\pm1 $ are correlated while the unsigned volumes $|\phi_i|$ are
i.i.d.\ and described by a generic distribution $p(|\phi_i|)$ defined on the finite positive support $(0,1)$. This  means that  the joint probability distribution
is factorizable as
\begin{equation}
P(\bphi_N)=\mathcal{P}(\bm{\epsilon}_N) \prod_{i=1}^N p(|\phi_i|).
\end{equation}
In this theoretical setup we are not able to compute analytically the price impact $I_N(\phi)$ and we will use numerical simulations.
To this aim we introduce a latent discrete variable $\tilde{\epsilon}$ in order to simulate  $N$ correlated signs  with  the following statistical model 
\begin{equation}
\mathbb{P}(\epsilon_i|\tilde{\epsilon})= \frac{1}{2} (1 +{\gamma}_\epsilon \epsilon_i \tilde \epsilon)
\label{eq:sampling_hidden}
\end{equation}
where $\gamma_\epsilon$ can be  estimated from data by averaging the realized sign correlation $\rho_\epsilon$ appearing in  Eq.~\eqref{eq:rhosign}, i.e.
\begin{equation}
\avg[\rho_\epsilon|N] = \gamma^2_\epsilon.
\label{eq:rhogammaeps}
\end{equation}
For clarity we explicitly omit the  $\gamma_\epsilon$'s dependence on $N$.
Thus to simulate the model we fix $\epsilon_k=+1$, we draw a hidden factor $\tilde \epsilon$ from
\begin{equation}
\label{eq:sampling_hidden2}
\mathbb{P}(\et|\epsilon_k=+1) = \frac{1}{2}(1+{\gamma}_\epsilon\et)
\end{equation}
and then we sample  $N-1$ other correlated signs $\{\epsilon_i\}_{i=1,\cdots,N}^{i \neq k}$ with probability
\begin{equation}\label{eq:pcond}
 \mathbb{P}(\epsilon_i|\tilde \epsilon)=\frac{1}{2}(1+\gamma_\epsilon \epsilon_i \tilde \epsilon) \, .
\end{equation}

\subsubsection{Numerical Computation of Market Impact}
\label{app:calibsign}
We summarize the main steps for the numerical calibration of the price impact $I_N(\phi)$ from data.
Given the number $N$ of metaorders per stock/day pair and fixing $|\phi_k|=\phi>0$:
\begin{enumerate}
\item We compute the average sign correlation $\mathbb{C}_\epsilon(N)=\avg[\rho_\epsilon|N]$ as to obtain $\gamma_\epsilon$ through Eq.~\eqref{eq:rhogammaeps}.
\item After fixing the direction $\epsilon_k= + 1$ we simulate $N-1$ correlated signs using Eq.~\eqref{eq:pcond}.
\item We sample $N-1$ random variables $|\phi_i|$ from an half-normal distribution with mean $\volfl_N \sqrt{2/\pi}$ and standard deviation $\volfl_N\sqrt{1-2/\pi}$  where $\volfl_N$ represents the empirical standard deviation of signed volumes.
\item We compute numerically the price impact $I_N(\phi)=\avg[\imp(\bphi_N)|\epsilon_k=+1,|\phi_k|=\phi]$ where  $\imp(\bphi_N) =  \left(\sum_{i=1}^N \phi_i\right)^{\bullet 1/2}$ and $\phi_i=\epsilon_i|\phi_i|$, as defined in Eq.~\eqref{eq:formula2}.
\item Finally, we compute $I(\phi)$  averaging $I_N(\phi)$ over the empirical distribution $p(N)$ shown in the left panel of Fig~\ref{figure7}. 
\end{enumerate}

\bibliographystyle{unsrt}

\end{document}